%% file: zh_llbb_prl.tex
\documentclass[prd,twocolumn,showpacs,amsmath,amssymb,superscriptaddress]{revtex4}

\usepackage{epsfig}
\usepackage{psfrag}
\usepackage{graphicx}
\usepackage{dcolumn}
\usepackage{bm}
\usepackage{epsfig}
\usepackage{lineno}
\usepackage{multirow}
\usepackage{slashed}
\usepackage{rotating}

\input{variables}

\input{definitions}

\def \et {E_{T}}
\def  \met {\not\!\!\et }

\begin{document}

\hspace{5.2in} \mbox{FERMILAB-PUB-12-403-E}





\title{\boldmath 
Search for the standard model Higgs boson 
in $ZH \rightarrow \ell^+\ell^-b\bar{b}$ production with the D0 detector in
$9.7$~fb$^{-1}$ of $p\bar{p}$ collisions at $\sqrt{s}=1.96~\text{TeV}$}



\input author_list.tex

\date{July 24, 2012}

\begin{abstract}
  We present a search for the standard model (SM) Higgs
  boson produced in association with a $Z$ boson in 9.7 fb$^{-1}$ of
  $p\bar{p}$ collisions collected with the \dzero\ detector at the
  Fermilab Tevatron Collider at $\sqrt{s}$ = 1.96 $\TeV$.  Selected events
  contain one reconstructed $Z\rightarrow e^+e^-$ or $Z\rightarrow
  \mu^+\mu^-$ candidate and at least two jets, including at least one
  jet identified as likely to contain a $b$ quark.  
  To validate the search procedure,
  we  also measure the cross section for $ZZ$ production 
  in the same final state.
  It is found to be consistent with its SM prediction. 
  We set upper limits on
  the $ZH$ production cross section times branching ratio for $H\to\bbbar$
  at the $95\%$ C.L. for
  Higgs boson masses $90 \leq M_H \leq 150~\GeV$. The observed (expected)
  limit for $M_H = 125~\GeV$ is 7.1 (5.1) times the SM cross section. 
\end{abstract}
\pacs{13.85.Ni, 13.85.Qk, 13.85.Rm, 14.80.Bn}

\maketitle

\newpage

In the standard model (SM), the spontaneous breaking of the
electroweak gauge symmetry generates masses for the $W$ and $Z$ bosons
and produces a residual massive particle, the Higgs boson~\cite{higgs}.  
Precision
electroweak data, including the latest $W$ boson mass measurements from the 
CDF \cite{cdf-wmass} and D0 \cite{d0-wmass} Collaborations, 
and the latest Tevatron combination for the top quark mass~\cite{tev-top}
constrain the mass of the 
SM Higgs boson to $M_H<$ 152 GeV \cite{ew-fit} at the 95\% 
confidence level (C.L.).  Direct
searches at the CERN $e^+e^-$ Collider (LEP)~\cite{lep-results}, by the
CDF and D0 Collaborations at the Fermilab Tevatron $\ppbar$
Collider~\cite{tev-results}, 
and by the ATLAS and CMS Collaborations
at the CERN Large Hadron Collider (LHC)~\cite{atlas-results,
  cms-results} further restrict the allowed range to $116.6 < M_{H} <
119.4$ GeV and $122.1 < M_{H} < 127.0$ GeV.  
The ATLAS and CMS results indicate excesses above
background expectations at $M_H~\approx~$125 GeV.
With additional data and analysis improvements,
the LHC experiments confirm their initial indications and
observe a particle with properties consistent with those
predicted for the SM Higgs boson~\cite{new-LHC-results}.

For ${M_H \lesssim 135~\GeV}$, the primary decay is
to the $b\bar{b}$ final state~\cite{hbr}.  
At the Tevatron, the best sensitivity
to a SM Higgs boson in this mass range is obtained from the analysis of its
production in association with a $W$ or $Z$
boson and its subsequent decay into $b\bar{b}$. 
Evidence for a signal in this decay mode would complement the
LHC findings and provide further indication that the new particle is the
SM Higgs boson.

We present a search for $ZH \rightarrow \ell^+\ell^-b\bar{b}$ events,
where $\ell$ is either a muon or an electron.  The data for this
analysis were collected at the Tevatron at $\sqrt{s}=$ 1.96 TeV with
the \dzero~detector from April 2002 to September 2011 
and correspond to an integrated luminosity of
9.7~fb$^{-1}$ after data quality requirements are imposed, which
represents the full Run II data set. To validate the search procedure,
we also present a measurement of the $ZZ$ production cross section in
the same final states and topologies used for the search.
The results presented here supersede our previous search in the $ZH
\rightarrow \ell^+\ell^-b\bar{b}$ channel~\cite{d0-zhiggs}.  
Beyond the inclusion of additional data, the most significant updates to
this analysis are the use of an improved $b$-jet identification algorithm,
revisions to the kinematic fit, and a new multivariate analysis strategy.
A search for $ZH\rightarrow \ell^+\ell^-b\bar{b}$ 
has also been performed by the CDF Collaboration~\cite{cdf-zhiggs}.

The D0 detector~\cite{d0det,d0upgrade} consists of a central tracking system
within a 2~T superconducting solenoidal magnet and surrounded by a
preshower detector, three liquid-argon sampling calorimeters, and a
muon spectrometer with a 1.8 T iron toroidal magnet.  
In the intercryostat regions (ICRs) between the central and end 
calorimeter cryostats, 
plastic scintillator detectors enhance the calorimeter coverage. 
The analyzed events were acquired predominantly with triggers that
select electron and muon candidates online.  
However, events
satisfying any trigger requirement are considered in this analysis.


The event selection requires a 
$\ppbar$
interaction vertex that has at least three associated tracks.
Selected events must contain a
$Z\to\ell^+\ell^-$ candidate. 
The analysis is conducted in four separate channels. 
The dimuon ($\mumu$) and dielectron ($\ee$) channels include 
events with at least two fully reconstructed muons or electrons.
In addition, muon-plus-track (\mumutrk) 
and electron-plus-ICR electron (\eeicr)
channels are designed to recover events in which one 
of the leptons points to
a poorly instrumented region of the detector.

The $\mumu$ event selection requires at least two muons identified
in the muon system, both matched to central tracks 
with transverse momenta $\pt >10~\GeV$.  
At least one muon must have $|\eta| < 1.5$,
where $\eta$ is the pseudorapidity, and $\pt > 15~\GeV$.
At least one of the muons
must be separated from any jet with $\pt>$ 20 GeV and $|\eta|<2.5$ by
$\Delta\mathcal{R} = \sqrt{\Delta\eta^2 + \Delta\phi^2} > 0.5$, 
from other tracks, and from energy deposited in the calorimeter.
We also apply isolation requirements based on the ratios
of the calorimeter energy and the sum of $\pt$ of tracks near the lepton to
the lepton $\pt$ in this analysis.

The $\mumutrk$ event selection
requires exactly one muon with $|\eta| < 1.5$ 
and $\pt > 15~\GeV$ that is isolated both in the tracker
and in the calorimeter.  
In addition, a second isolated track reconstructed in the tracker
with $|\eta| < 2$ and $\pt > 20~\GeV$ must be present.  
Its distance $\Delta\mathcal{R} $ from the muon 
and from any jet of $\pt >15~\GeV$ and  $|\eta| < 2.5$ must be greater
than 0.1 and 0.5, respectively.
For the $\mumu$ and $\mumutrk$ channels, the two muon-associated tracks must 
have opposite charge.

The $\ee$ event selection requires at least two electrons with transverse
energy $\et >
15~\GeV$ that pass selection requirements based on 
the energy deposition and shower shape in the calorimeter 
and the preshower detector.
Both electrons are required to be isolated in the tracker 
and the calorimeter.
At least one electron must be identified
in the region $|\eta| < 1.1$.
The electrons in $|\eta| < 1.1$ must match central tracks or 
a set of hits in the tracker consistent with that of an electron trajectory. 

The $\eeicr$ event selection requires exactly one electron in the calorimeter
with $\et>15~\GeV$ and a track pointing toward one of the 
ICRs, $1.1<|\eta|<1.5$.
The track must be isolated, be matched to a calorimeter energy deposit with
$\et>10~\GeV$ and have $\pt>15~\GeV$.
For the
\ee~and \eeicr~selections, 
electrons must be separated from all jets by 
$\Delta\mathcal{R}  > 0.5$.

Jets are reconstructed in the calorimeter by using the iterative midpoint
cone algorithm~\cite{runiicone} with a cone of radius
0.5 in rapidity and azimuthal angle.
The jet identification efficiency is ${\approx 95\%}$ 
at $\pt = 20$~GeV and reaches 99\% at $\pt = 50$~GeV.
Jets are denoted as ``taggable'' if the associated tracks meet criteria 
that algorithms to identify jets as likely to contain $b$-quarks operate 
efficiently.
The taggability efficiency is at least 90\%
for most of the jets in 
this analysis.
We use ``inclusive'' to denote the event sample selected by
requiring the presence of two leptons and use ``pretag'' for the event
sample that meets the additional requirements of having at least
two taggable jets with $\pt > 20~\GeV$ and $|\eta| < 2.5$ 
and a dilepton invariant mass $70 < m_{\ell \ell} < 110~\GeV$~\cite{appendix}. 

Jets are
identified as likely to contain $b$ quarks ($b$-tagged) if they pass
``loose'' or ``tight'' requirements on the output of a multivariate
discriminant trained to separate $b$ jets from light jets.
This
discriminant is an improved version of the neural network $b$-tagging
discriminant described in Ref.~\cite{bid}.
For taggable jets in $|\eta|<1.1$ and with
$\pt\approx50~\GeV$, the $b$-tagging efficiency for $b$ jets and the
misidentification probability of light ($uds$ or gluon) jets are, 
respectively, $72\%$ and
$6.7\%$ for loose $b$ tags, and $47\%$ and $0.4\%$ for tight $b$ tags.
Events with at least one tight and one loose $b$ tag are classified as
double-tagged (DT).  Events not in the DT sample that contain a single
tight $b$ tag are classified as single-tagged (ST).  

The dominant background process is the production of a $Z$ boson in
association with jets, with the $Z$ decaying to dileptons ($Z+$jets).
The light-flavor component ($Z+$LF) includes jets from only light
quarks or gluons.  The heavy-flavor component ($Z+$HF)
includes $Z+b\bar{b}$, which has the same final state as
the signal, and $Z+c\bar{c}$ production.  The remaining
backgrounds are from $t\bar{t}$ production; $WW$, $WZ$, and $ZZ$ (diboson) 
production; and multijet (MJ) events with nonprompt muons 
or with jets misidentified as electrons.  

We simulate $ZH$ and diboson production with
\pythia~\cite{pythia}.  In the $ZH$ samples, we consider the
contributions to the signal from the 
$\ell^+\ell^-b{\bar b}$, $\ell^+\ell^-c{\bar c}$, and
$\ell^+\ell^-\tau^+\tau^-$ final states.  
The $\ell^+\ell^-\bbbar$ accounts for 99\% (97\%) of the signal yield in the DT
(ST) sample.
The $Z+$jets and 
$t\bar{t}$
processes are simulated
with \alpgen~\cite{alpgen}, followed by \pythia~for parton showering and
hadronization~\cite{mlm}.
All simulated samples are generated by using the {\sc CTEQ6L1}~\cite{cteq6} 
leading-order parton distribution functions.
We process all samples by using a detector
simulation program based on \geant~\cite{geant} and the same off\-line
reconstruction algorithms used for data. 
We overlay events from randomly
chosen beam crossings 
with the same instantaneous luminosity distribution as data
on the generated events to model the
effects of multiple $\ppbar$ interactions and detector noise.

\begin{table*}[htbp!]
\begin{centering}
\caption{
Expected and observed event yields for all lepton channels combined
after requiring two leptons (inclusive), after also requiring at least two jets
(pretag), and after requiring exactly one (ST) or at least two (DT)
$b$-tags.  
The $ZH$ signal yields are for $M_H=125$ GeV. 
The uncertainties quoted on the total background for ST and DT and signal include
the statistical and systematic uncertainties.
}
\begin{tabular}{lcccccccrcl}
\hline\hline
& Data
& Total background
& MJ
& $Z+$LF
& $Z+$HF
& Diboson
& $\ttbar$
& \multicolumn{3}{c}{$ZH$} \\
\hline
 Inclusive  &$1845610$ & $1841683$        & $160746$  & $1630391$  & $46462$ & $2914$ & $1170$ & $17.3$&$\pm$&$1.1$  \\
 Pretag     &    $25849$ & $25658$       &   $1284$   &    $19253$   &   $4305$ &   $530$ &   $285$ & $9.2$&$\pm$&$0.6$ \\
 ST         &       $886$ &       $824\pm102$ &      $54$   &        $60$   &     $600$ &    $33$ &    $77$ & $2.5$&$\pm$&$0.2$  \\
 DT         &       $373$ &       $366\pm39$  &      $25.7$ &         $3.5$ &     $219$ &    $19$ &    $99$ & $2.9$&$\pm$&$0.2$  \\
\hline\hline
\end{tabular}
\label{tbl:evtall}
\end{centering}
\end{table*}

We take the cross sections and branching ratios for signal from 
Refs.~\cite{hbr,zhxsec}.  For 
the diboson processes, we use
next-to-leading order (NLO) cross sections from the Monte Carlo program 
\mcfm~\cite{mcfm}.
We scale the $\ttbar$ cross section to approximate 
next-to-NLO~\cite{ttbarxsec} 
and the inclusive $Z$ boson cross section
to next-to-NLO~\cite{dyxsec} and apply additional NLO heavy-flavor
corrections to the $Z+b\bar{b}$ and $Z+c\bar{c}$ samples, 
calculated from \mcfm\ to be 1.52 and 1.67, respectively.

To improve the modeling of the $\pt$ distribution of the $Z$ boson, 
we reweight simulated $Z$+jets events to be consistent with
the measured $\pt$ spectrum of  $Z$ bosons in the data~\cite{zptrw}.
We correct the energies of simulated jets to reproduce the
resolution and energy scale observed in the data~\cite{jetescale}. 
We apply the trigger efficiencies, measured in the data, as event
weights to the simulated $\mumu$, $\mumutrk$ and $\eeicr$ events.
In the $ee$ channel, we have verified that the trigger efficiency is 
consistent with 100\% for our selection.
We apply scale factors to account for differences in reconstruction efficiency
between the data and simulation.  
Motivated by a comparison with data~\cite{zjets}
and the {\sc sherpa} generator~\cite{sherpa},
we reweight the $Z+$jets events to improve the {\alpgen} modeling 
of the distributions of the $\eta$ of the two
jets.

We estimate the MJ backgrounds from control samples in
data obtained by inverting some of the lepton 
selection requirements, e.g., the lepton isolation requirements in the
$\mumu$ channel and the shower shape requirements in the $\ee$ channel.
%
We adjust the normalizations of the MJ background and all simulated
samples by scale factors
determined from a simultaneous fit to the $\mll$ distributions in the
0-jet, 1-jet, and $\geq$ 2-jet samples of each lepton selection.  
The inclusive sample
constrains the lepton trigger and identification efficiencies, while
the pretag sample, which includes jet requirements, is used to correct
the $Z+$jets cross section.
The total event yields
after applying all corrections and normalization factors are shown in
Table \ref{tbl:evtall}.  
The observed event yields are consistent with the expected background.

To exploit the fully constrained kinematics of 
the $ZH \rightarrow \ell^+\ell^- \bbbar$
process, we adjust the energies of the candidate leptons and jets
within their experimental resolutions by using a likelihood fit that
constrains $m_{\ell\ell}$ to the mass and width of the $Z$ boson and constrains
the $\pt$ of the $\ell^+\ell^-\bbbar$ system to zero with an expected width 
determined from $ZH$ Monte Carlo events. 
This kinematic fit improves the dijet mass resolution
by 10\%$-$15\%, depending on $M_H$.
The dijet mass resolution for $M_H=125$~GeV is $\approx 15$~GeV with the
kinematic fit~\cite{appendix}.

We use a two step multivariate analysis strategy based on random forest 
(RF, an ensemble classifier that consists of many decision trees) 
discriminants~\cite{dtree}, as implemented in the {\sc tmva} 
software package~\cite{tmva},
to improve the separation of signal from background~\cite{appendix}.
We choose well modeled kinematic variables that are sensitive to 
the $ZH$ signal as inputs for the analysis. These include
the 
$\pt$ of the two $b$-jet candidates and the dijet mass, 
before and after the jet energies are adjusted by the kinematic fit.
In the first step, we train a dedicated RF ($\ttbar$ RF) that takes
$\ttbar$ as the only background and $ZH$ as the signal.  This approach
takes advantage of the characteristic signature of the $\ttbar$ background, for
instance, the presence of large missing transverse energy.
In the second step, we use the
$t\bar{t}$ RF to define two independent regions: a $t\bar{t}$ enriched
region ($t\bar{t}$ RF $<0.5$) and a $t\bar{t}$ depleted region
($t\bar{t}$ RF $\ge 0.5$).  
The $\ttbar$ depleted region contains 94\% (93\%) of the DT (ST) signal
contribution and 55\% (82\%) of DT (ST) background events. 
In each region, we train a global RF to
separate the $ZH$ signal from all backgrounds.  In both steps we consider
ST and DT events separately and train the discriminants for each
assumed value of $M_H$ in 5~GeV steps from 90 to 150~GeV.

We assess systematic uncertainties resulting from the background normalization
for the MJ contribution, typically $10\%$.
The normalization of the $Z+$jets sample to the pretag
data constrains that sample to the statistical uncertainty, $<$1\%, of the
pretag data. Because this sample is dominated by the $Z$+LF background,
the normalization of the $\ttbar$, diboson, and $ZH$ samples
acquires a sensitivity to the inclusive $Z$ cross section, for which
we assess a 6\% uncertainty \cite{dyxsec}. We assign this uncertainty to
these samples as a common uncertainty.
For $Z+$HF, a cross
section uncertainty of 20\% is determined from Ref.~\cite{mcfm}.  For
other backgrounds, the uncertainties are 6\%--10\%  \cite{mcfm,ttbarxsec}. 
For the
signal, the cross section uncertainty is 6\%~\cite{zhxsec}. 
Sources of systematic uncertainty affecting the shapes of the 
final discriminant distributions are
the jet energy scale, 1\%--3\%; 
jet energy resolution, 2\%--4\%; 
jet identification efficiency, $\approx 4\%$;
and $b$-tagging efficiency, 4\%--6\%.
Other sources include trigger efficiency, 4\%--6\%; parton distribution function 
uncertainties~\cite{pdf}, $<$1\%;
data-determined corrections to the model for $Z+$jets, 3\%--4\%;
modeling of the underlying event, $<$1\%;
and from varying the factorization and
renormalization scales for the $Z+$jets simulation, $<$1\%.

\begin{figure}[htbp]
\includegraphics[height=0.5\textheight]{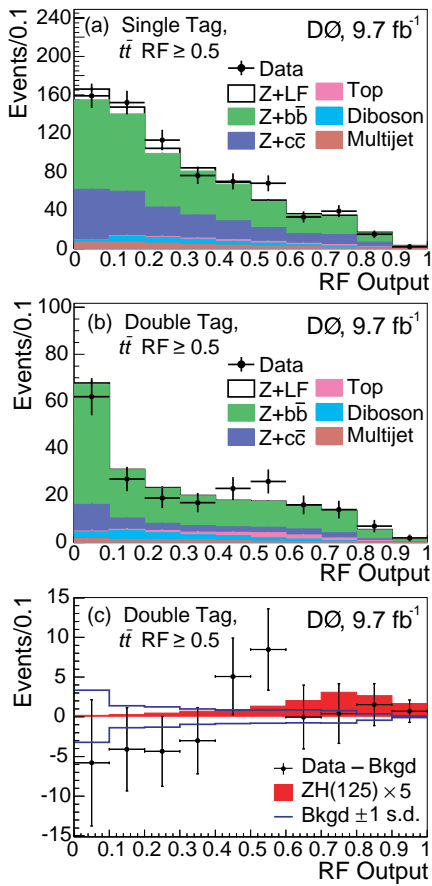}
\caption{\label{fig:rf_poor_postfit} (color online).
Distributions of the global RF discriminant
in the $\ttbar$ depleted region, assuming $M_H=$ 125 GeV, 
after the fit to the background-only model
for data (points with statistical error bars) and background (histograms) 
for (a) single-tagged events and (b) double-tagged events.  
(c) Background-subtracted distribution for (b).
The signal distribution is shown with the SM cross section 
scaled by a factor of five. The blue lines 
indicate the uncertainty from the fit.}
\end{figure}

The global RF distributions from the four samples (ST and DT in the
$\ttbar$ depleted and $\ttbar$ enriched regions) in each channel along
with the corresponding systematic uncertainties are used for the
statistical analysis of the data.  We set 95\% C.L.~upper limits on the
$ZH$ cross section times branching ratio for $H\to\bbbar$ with a
modified frequentist (CL$_s$) method that uses the log likelihood
ratio of the signal+background (S+B) hypothesis to the background-only
(B) hypothesis \cite{cls}. 
To minimize the effect of
systematic uncertainties, we maximize the likelihoods of the B and S+B 
hypotheses by independent fits that allow the sources of 
systematic uncertainty to vary within their Gaussian
priors~\cite{wade}.

To validate the search procedure, we search for $ZZ$ production in 
the $\ell^+\ell^- \bbbar$ and $\ell^+\ell^- \ccbar$ final states.
We use the same event selection, corrections to our signal and background
models, and RF training procedure as for the $ZH$ 
search~\cite{appendix}.
Our search also includes
$WZ$ production in the $\csbar\ell^+\ell^-$ final state.
We collectively refer to these as $VZ$ production.
Using the same modified frequentist method
as for the $ZH$ search and fitting the RF distributions to the S+B
hypothesis,
we measure a $VZ$ cross section of 
$0.8 \pm 0.4\thinspace({\rm stat}) \pm 0.4\thinspace({\rm syst})$ times 
that of the SM prediction
with a significance of $1.5$ standard deviation (s.d.) and an expected 
significance of $1.9$ s.d.
This result
is consistent with the recent D0 $ZZ+WZ$ cross section measurement 
obtained in fully leptonic decay channels~\cite{d0zzwz}.

\begin{figure}[htbp!]
\includegraphics[height=0.22\textheight]{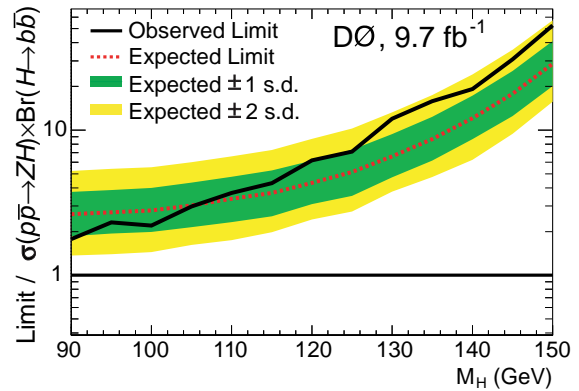}
\caption{(color online). Expected and observed 95\% C.L. cross section 
upper limits on the
$ZH$ cross section times branching ratio for $H\to\bbbar$,
  expressed as a ratio to the SM prediction.
}
\label{fig:limits}
\end{figure}

The output of the RF trained to separate signal events with $M_H=$ 125 GeV
from background is shown 
in Fig.~\ref{fig:rf_poor_postfit} for ST and DT events separately in the
$\ttbar$ depleted region, after the
background-only fit.  
Also shown is the background-subtracted 
RF distribution for DT events in the data.
The upper limit on the cross section times branching ratio  for
$H\to\bbbar$, expressed as a ratio to the SM prediction, is presented 
as a function of $M_H$ in
Table~\ref{tbl:limits} and 
Fig.~\ref{fig:limits}.  At $M_H=125~\GeV$,
the observed (expected) limit on this ratio is 7.1 (5.1).
The expected limits are $\approx 20\%$ lower than those anticipated 
from the increase in the data because of the analysis improvements
described above.

\begin{table*}[htbp!]
\caption{The expected and observed 95\% C.L. upper limits on the $ZH$ 
production cross section times branching ratio for $ZH \rightarrow \ell^+\ell^- b\bar{b}$,
expressed as a ratio to the SM prediction.}
\begin{center}
\begin{tabular}{lccccccccccccc}
\hline
 $M_H~(\GeV)$ & 90 & 95 & 100 & 105 & 110 & 115 & 120 & 125 & 130 & 135 & 140 & 145 & 150 \\
\hline\hline
Expected & 2.6 & 2.7 & 2.8 & 3.0 & 3.4 & 3.7 & 4.3 & 5.1 & 6.6 & 8.7 & 12 & 18 & 29\\
Observed & 1.8 & 2.3 & 2.2 & 3.0 & 3.7 & 4.3 & 6.2 & 7.1 & 12 & 16 & 19 & 31 & 53\\
\hline
\end{tabular}
\label{tbl:limits}
\end{center}
\end{table*}

In summary, we have searched for SM Higgs boson production in association
with a $Z$ boson in the final state of two charged leptons (electrons
or muons) and two $b$-quark jets by using a 9.7 fb$^{-1}$ data set of
$p\bar{p}$ collisions at $\sqrt{s}$ = 1.96 $\TeV$. 
We also measure the cross section for $VZ$ production 
in the same final state with the result of 
$0.8 \pm 0.4\thinspace({\rm stat}) \pm 0.4\thinspace({\rm syst})$ 
times its SM prediction.
We set an upper limit on the $ZH$ production cross section times
branching ratio for $H\to\bbbar$ as a function of $M_H$.  
The observed (expected) limit for 
$M_H = 125~\GeV$ is 7.1 (5.1) times the SM 
cross section.  

\begin{acknowledgments}
\input{acknowledgement}
\end{acknowledgments}

\input{epaps}

\end{document}

%% file: variables.tex
\newcommand{\pt}{\mbox{$p_{T}$}}
\newcommand{\et}{\mbox{$E_{T}$}}
\newcommand{\met}{\mbox{$\slashed{E}_{T}$}}
\newcommand{\metsig}{\mbox{$\slashed{E}_{T}^{sig}$}}

\newcommand{\mbb}{\mbox{$m_{bb}$}}
\newcommand{\ptbb}{\mbox{$p_{T}^{bb}$}}
\newcommand{\ptbone}{\mbox{$p_{T}^{b1}$}}
\newcommand{\ptbtwo}{\mbox{$p_{T}^{b2}$}}
\newcommand{\dphibb}{\mbox{$\Delta\phi(b_1,b_2)$}}
\newcommand{\detabb}{\mbox{$\Delta\eta(b_1,b_2)$}}

\newcommand{\ptz}{\mbox{$p_{T}^{Z}$}}
\newcommand{\dphill}{\mbox{$\Delta\phi(\ell_1,\ell_2)$}}

\newcommand{\colll}{\mbox{${\rm colinearity}(\ell_1, \ell_2)$}}
\newcommand{\dphizbb}{\mbox{$\Delta\phi(\ell\ell,bb)$}}
\newcommand{\costhst}{\mbox{$\cos\theta^*$}}
\newcommand{\mllbb}{\mbox{$m(\ell \ell bb)$}}
\newcommand{\htllbb}{\mbox{$H_T(\ell \ell bb)$}}
\newcommand{\mjets}{\mbox{$m(\sum{j_i})$}}
\newcommand{\ptjets}{\mbox{$\pt(\sum{j_i})$}}
\newcommand{\htjets}{\mbox{$H_T(\sum{j_i})$}}


%% file: definitions.tex
\newcommand{\dzero}{D0}

\newcommand{\ee}      {\ensuremath{ee}}
\newcommand{\mumu}    {\ensuremath{\mu\mu}}
\newcommand{\eeicr}   {\ensuremath{ee_{\rm ICR}}}
\newcommand{\mumutrk} {\ensuremath{\mu\mu_{\rm trk}}}

\newcommand{\GeV} {\ensuremath{\mathrm{Ge\kern -0.1em V}}}
\newcommand{\TeV} {\ensuremath{\mathrm{Te\kern -0.1em V}}}

\newcommand{\ppbar}{\mbox{$p\overline{p}$}}

\newcommand{\ccbar}{\mbox{$c\overline{c}$}}
\newcommand{\bbbar}{\mbox{$b\overline{b}$}}
\newcommand{\ttbar}{\mbox{$t\overline{t}$}}
\newcommand{\csbar}{\mbox{$c\overline{s}$}}

\newcommand{\alpgen}{{\sc alpgen}}
\newcommand{\pythia}{{\sc pythia}}
\newcommand{\mcfm}{{\sc mcfm}}
\newcommand{\geant}{{\sc geant3}}

\newcommand{\mll}{\mbox{$m_{\ell \ell}$}}

%% file: author_list.tex
%
\affiliation{LAFEX, Centro Brasileiro de Pesquisas F\'{i}sicas, Rio de Janeiro, Brazil}
\affiliation{Universidade do Estado do Rio de Janeiro, Rio de Janeiro, Brazil}
\affiliation{Universidade Federal do ABC, Santo Andr\'e, Brazil}
\affiliation{University of Science and Technology of China, Hefei, People's Republic of China}
\affiliation{Universidad de los Andes, Bogot\'a, Colombia}
\affiliation{Charles University, Faculty of Mathematics and Physics, Center for Particle Physics, Prague, Czech Republic}
\affiliation{Czech Technical University in Prague, Prague, Czech Republic}
\affiliation{Center for Particle Physics, Institute of Physics, Academy of Sciences of the Czech Republic, Prague, Czech Republic}
\affiliation{Universidad San Francisco de Quito, Quito, Ecuador}
\affiliation{LPC, Universit\'e Blaise Pascal, CNRS/IN2P3, Clermont, France}
\affiliation{LPSC, Universit\'e Joseph Fourier Grenoble 1, CNRS/IN2P3, Institut National Polytechnique de Grenoble, Grenoble, France}
\affiliation{CPPM, Aix-Marseille Universit\'e, CNRS/IN2P3, Marseille, France}
\affiliation{LAL, Universit\'e Paris-Sud, CNRS/IN2P3, Orsay, France}
\affiliation{LPNHE, Universit\'es Paris VI and VII, CNRS/IN2P3, Paris, France}
\affiliation{CEA, Irfu, SPP, Saclay, France}
\affiliation{IPHC, Universit\'e de Strasbourg, CNRS/IN2P3, Strasbourg, France}
\affiliation{IPNL, Universit\'e Lyon 1, CNRS/IN2P3, Villeurbanne, France and Universit\'e de Lyon, Lyon, France}
\affiliation{III. Physikalisches Institut A, RWTH Aachen University, Aachen, Germany}
\affiliation{Physikalisches Institut, Universit\"at Freiburg, Freiburg, Germany}
\affiliation{II. Physikalisches Institut, Georg-August-Universit\"at G\"ottingen, G\"ottingen, Germany}
\affiliation{Institut f\"ur Physik, Universit\"at Mainz, Mainz, Germany}
\affiliation{Ludwig-Maximilians-Universit\"at M\"unchen, M\"unchen, Germany}
\affiliation{Fachbereich Physik, Bergische Universit\"at Wuppertal, Wuppertal, Germany}
\affiliation{Panjab University, Chandigarh, India}
\affiliation{Delhi University, Delhi, India}
\affiliation{Tata Institute of Fundamental Research, Mumbai, India}
\affiliation{University College Dublin, Dublin, Ireland}
\affiliation{Korea Detector Laboratory, Korea University, Seoul, Korea}
\affiliation{CINVESTAV, Mexico City, Mexico}
\affiliation{Nikhef, Science Park, Amsterdam, the Netherlands}
\affiliation{Radboud University Nijmegen, Nijmegen, the Netherlands}
\affiliation{Joint Institute for Nuclear Research, Dubna, Russia}
\affiliation{Institute for Theoretical and Experimental Physics, Moscow, Russia}
\affiliation{Moscow State University, Moscow, Russia}
\affiliation{Institute for High Energy Physics, Protvino, Russia}
\affiliation{Petersburg Nuclear Physics Institute, St. Petersburg, Russia}
\affiliation{Instituci\'{o} Catalana de Recerca i Estudis Avan\c{c}ats (ICREA) and Institut de F\'{i}sica d'Altes Energies (IFAE), Barcelona, Spain}
\affiliation{Uppsala University, Uppsala, Sweden}
\affiliation{Lancaster University, Lancaster LA1 4YB, United Kingdom}
\affiliation{Imperial College London, London SW7 2AZ, United Kingdom}
\affiliation{The University of Manchester, Manchester M13 9PL, United Kingdom}
\affiliation{University of Arizona, Tucson, Arizona 85721, USA}
\affiliation{University of California Riverside, Riverside, California 92521, USA}
\affiliation{Florida State University, Tallahassee, Florida 32306, USA}
\affiliation{Fermi National Accelerator Laboratory, Batavia, Illinois 60510, USA}
\affiliation{University of Illinois at Chicago, Chicago, Illinois 60607, USA}
\affiliation{Northern Illinois University, DeKalb, Illinois 60115, USA}
\affiliation{Northwestern University, Evanston, Illinois 60208, USA}
\affiliation{Indiana University, Bloomington, Indiana 47405, USA}
\affiliation{Purdue University Calumet, Hammond, Indiana 46323, USA}
\affiliation{University of Notre Dame, Notre Dame, Indiana 46556, USA}
\affiliation{Iowa State University, Ames, Iowa 50011, USA}
\affiliation{University of Kansas, Lawrence, Kansas 66045, USA}
\affiliation{Kansas State University, Manhattan, Kansas 66506, USA}
\affiliation{Louisiana Tech University, Ruston, Louisiana 71272, USA}
\affiliation{Boston University, Boston, Massachusetts 02215, USA}
\affiliation{Northeastern University, Boston, Massachusetts 02115, USA}
\affiliation{University of Michigan, Ann Arbor, Michigan 48109, USA}
\affiliation{Michigan State University, East Lansing, Michigan 48824, USA}
\affiliation{University of Mississippi, University, Mississippi 38677, USA}
\affiliation{University of Nebraska, Lincoln, Nebraska 68588, USA}
\affiliation{Rutgers University, Piscataway, New Jersey 08855, USA}
\affiliation{Princeton University, Princeton, New Jersey 08544, USA}
\affiliation{State University of New York, Buffalo, New York 14260, USA}
\affiliation{University of Rochester, Rochester, New York 14627, USA}
\affiliation{State University of New York, Stony Brook, New York 11794, USA}
\affiliation{Brookhaven National Laboratory, Upton, New York 11973, USA}
\affiliation{Langston University, Langston, Oklahoma 73050, USA}
\affiliation{University of Oklahoma, Norman, Oklahoma 73019, USA}
\affiliation{Oklahoma State University, Stillwater, Oklahoma 74078, USA}
\affiliation{Brown University, Providence, Rhode Island 02912, USA}
\affiliation{University of Texas, Arlington, Texas 76019, USA}
\affiliation{Southern Methodist University, Dallas, Texas 75275, USA}
\affiliation{Rice University, Houston, Texas 77005, USA}
\affiliation{University of Virginia, Charlottesville, Virginia 22904, USA}
\affiliation{University of Washington, Seattle, Washington 98195, USA}
\author{V.M.~Abazov} \affiliation{Joint Institute for Nuclear Research, Dubna, Russia}
\author{B.~Abbott} \affiliation{University of Oklahoma, Norman, Oklahoma 73019, USA}
\author{B.S.~Acharya} \affiliation{Tata Institute of Fundamental Research, Mumbai, India}
\author{M.~Adams} \affiliation{University of Illinois at Chicago, Chicago, Illinois 60607, USA}
\author{T.~Adams} \affiliation{Florida State University, Tallahassee, Florida 32306, USA}
\author{G.D.~Alexeev} \affiliation{Joint Institute for Nuclear Research, Dubna, Russia}
\author{G.~Alkhazov} \affiliation{Petersburg Nuclear Physics Institute, St. Petersburg, Russia}
\author{A.~Alton$^{a}$} \affiliation{University of Michigan, Ann Arbor, Michigan 48109, USA}
\author{G.~Alverson} \affiliation{Northeastern University, Boston, Massachusetts 02115, USA}
\author{A.~Askew} \affiliation{Florida State University, Tallahassee, Florida 32306, USA}
% ...
%
%
\collaboration{The D0 Collaboration\footnote{with visitors from
$^{a}$Augustana College, Sioux Falls, SD, USA,
$^{b}$The University of Liverpool, Liverpool, UK,
$^{c}$UPIITA-IPN, Mexico City, Mexico,
$^{d}$DESY, Hamburg, Germany,
$^{e}$SLAC, Menlo Park, CA, USA,
$^{f}$University College London, London, UK,
$^{g}$Centro de Investigacion en Computacion - IPN, Mexico City, Mexico,
$^{h}$ECFM, Universidad Autonoma de Sinaloa, Culiac\'an, Mexico
and
$^{i}$Universidade Estadual Paulista, S\~ao Paulo, Brazil.
}} \noaffiliation
\vskip 0.25cm

%% file: acknowledgement.tex
%
We thank the staffs at Fermilab and collaborating institutions,
and acknowledge support from the
DOE and NSF (USA);
CEA and CNRS/IN2P3 (France);
FASI, Rosatom and RFBR (Russia);
CNPq, FAPERJ, FAPESP and FUNDUNESP (Brazil);
DAE and DST (India);
Colciencias (Colombia);
CONACyT (Mexico);
KRF and KOSEF (Korea);
CONICET and UBACyT (Argentina);
FOM (The Netherlands);
STFC and the Royal Society (United Kingdom);
MSMT and GACR (Czech Republic);
CRC Program and NSERC (Canada);
BMBF and DFG (Germany);
SFI (Ireland);
The Swedish Research Council (Sweden);
and
CAS and CNSF (China).

%% file: epaps.tex




\hyphenation{ALPGEN}



\newpage

\begin{widetext}


\appendix*
\section{}


\begin{figure*}[htbp!]
\begin{centering}
\begin{tabular}{cc}
\includegraphics[height=0.18\textheight]{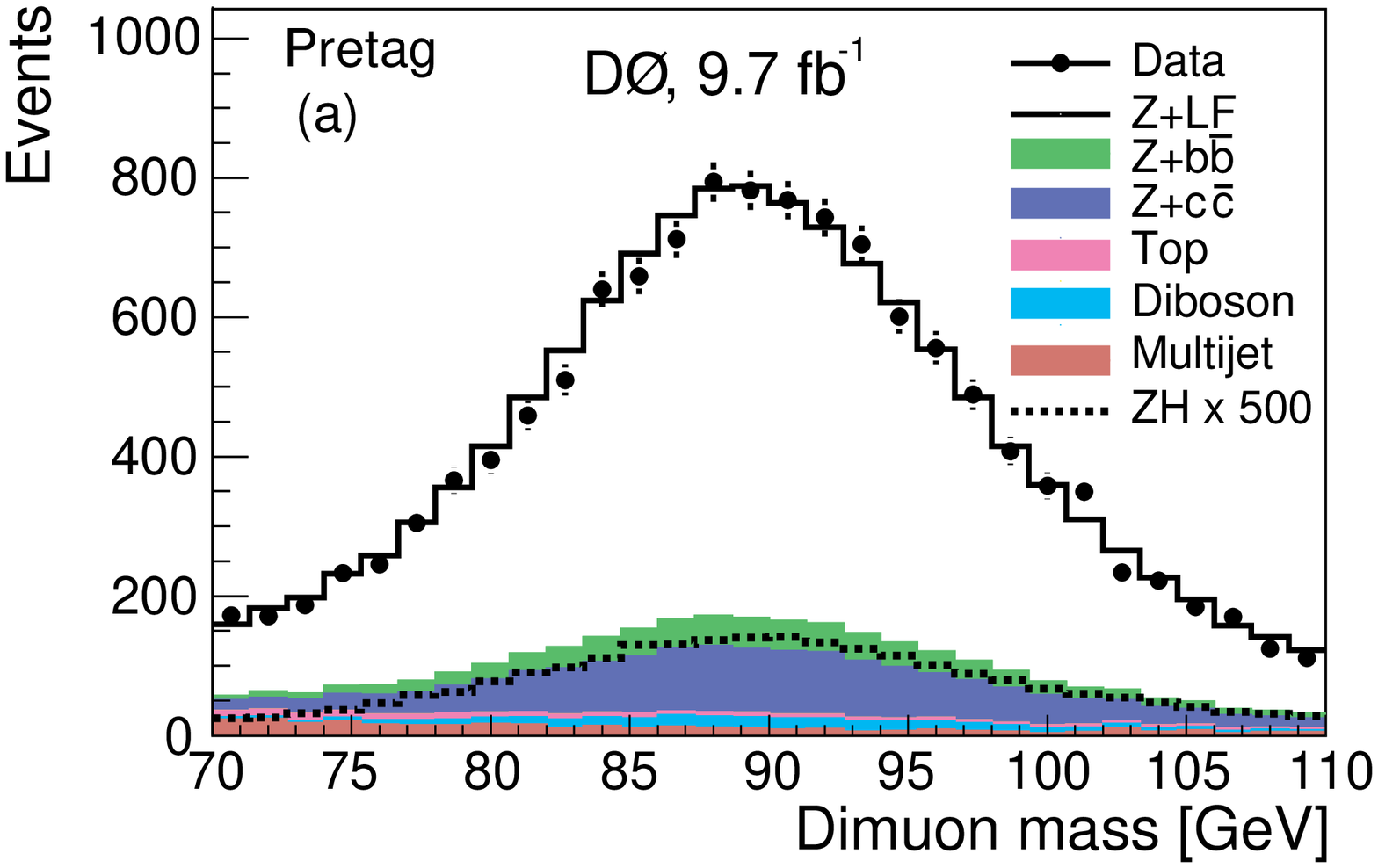} & 
\includegraphics[height=0.18\textheight]{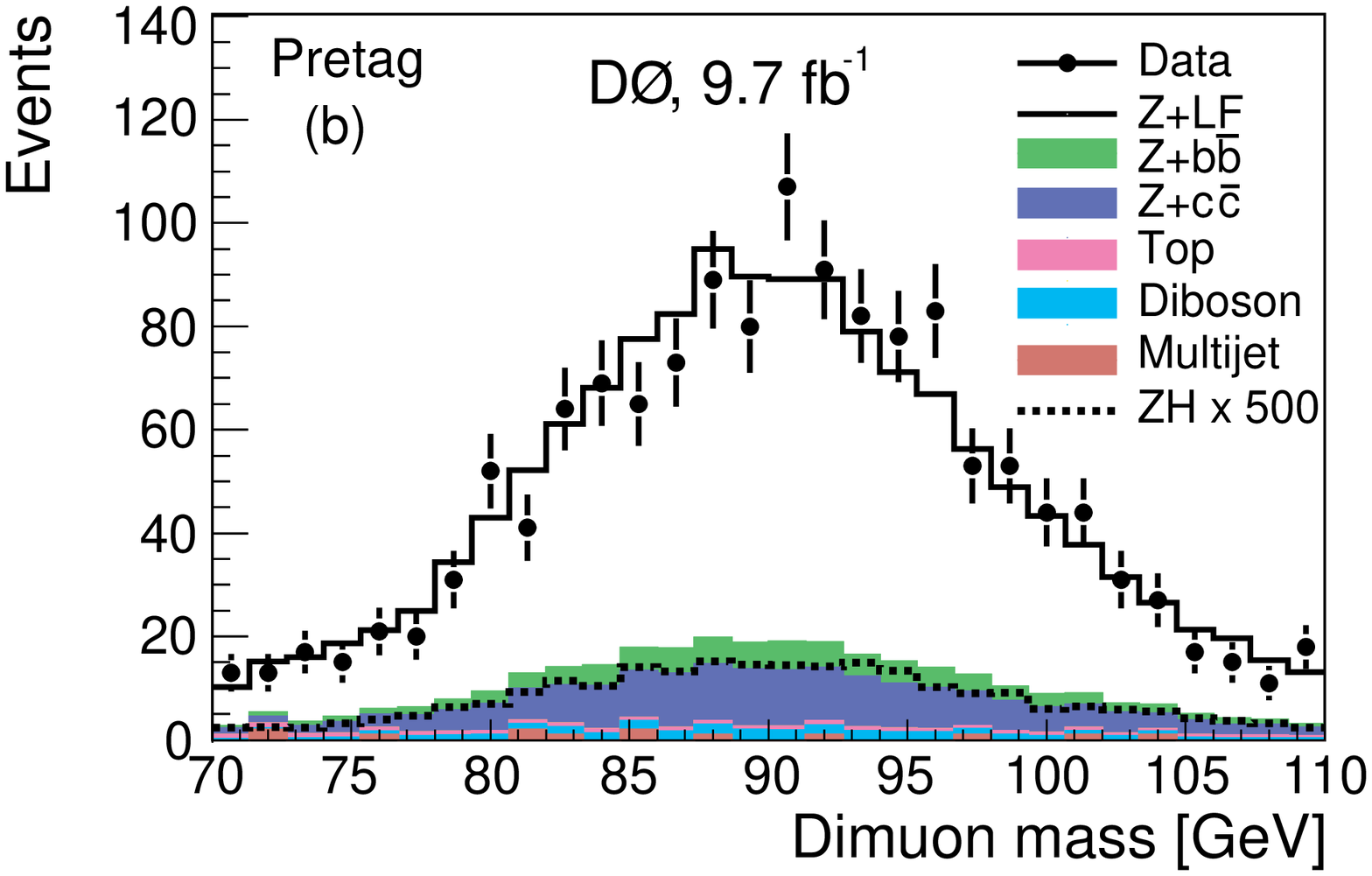} \\  
\includegraphics[height=0.18\textheight]{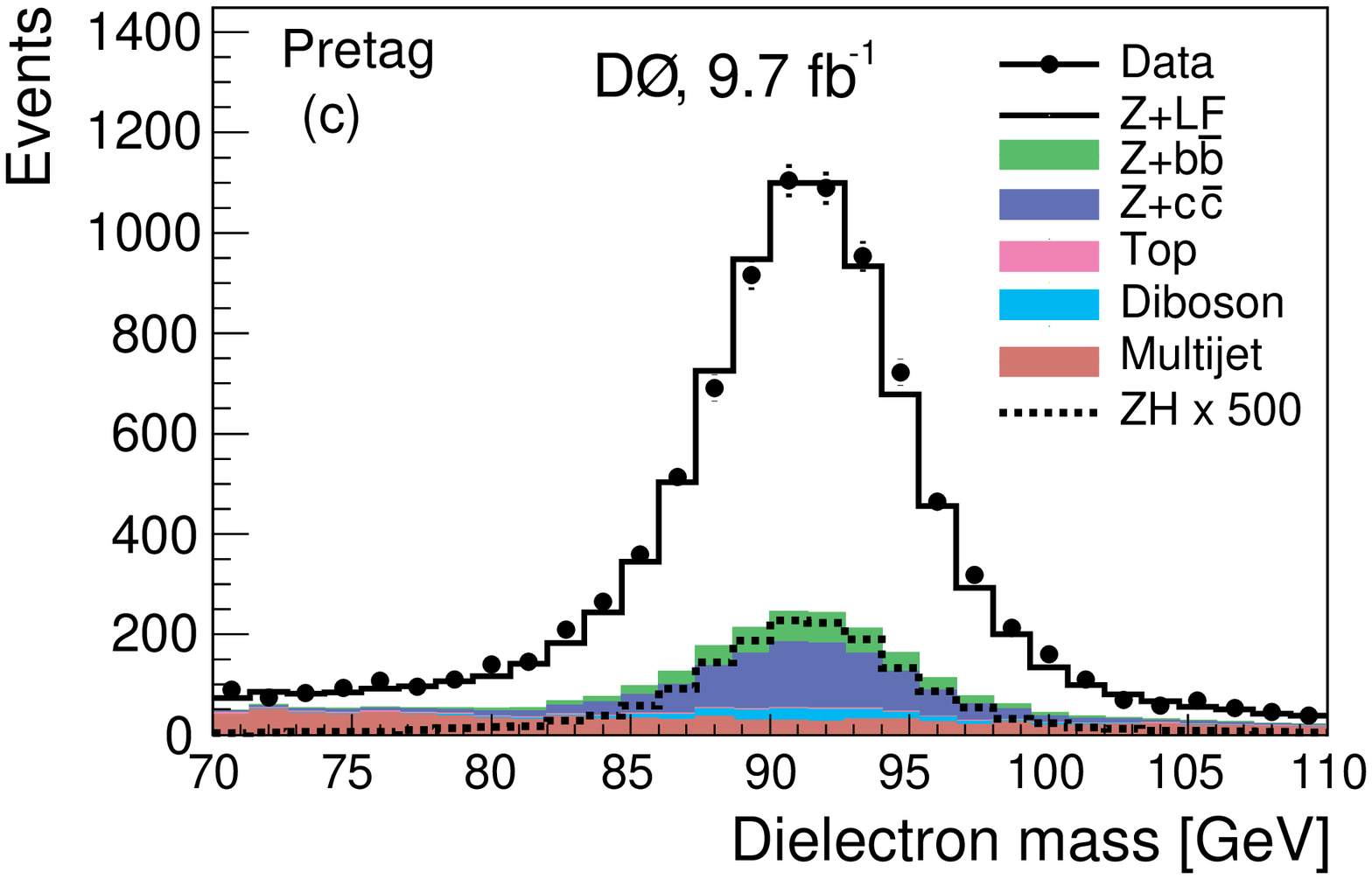} &
\includegraphics[height=0.18\textheight]{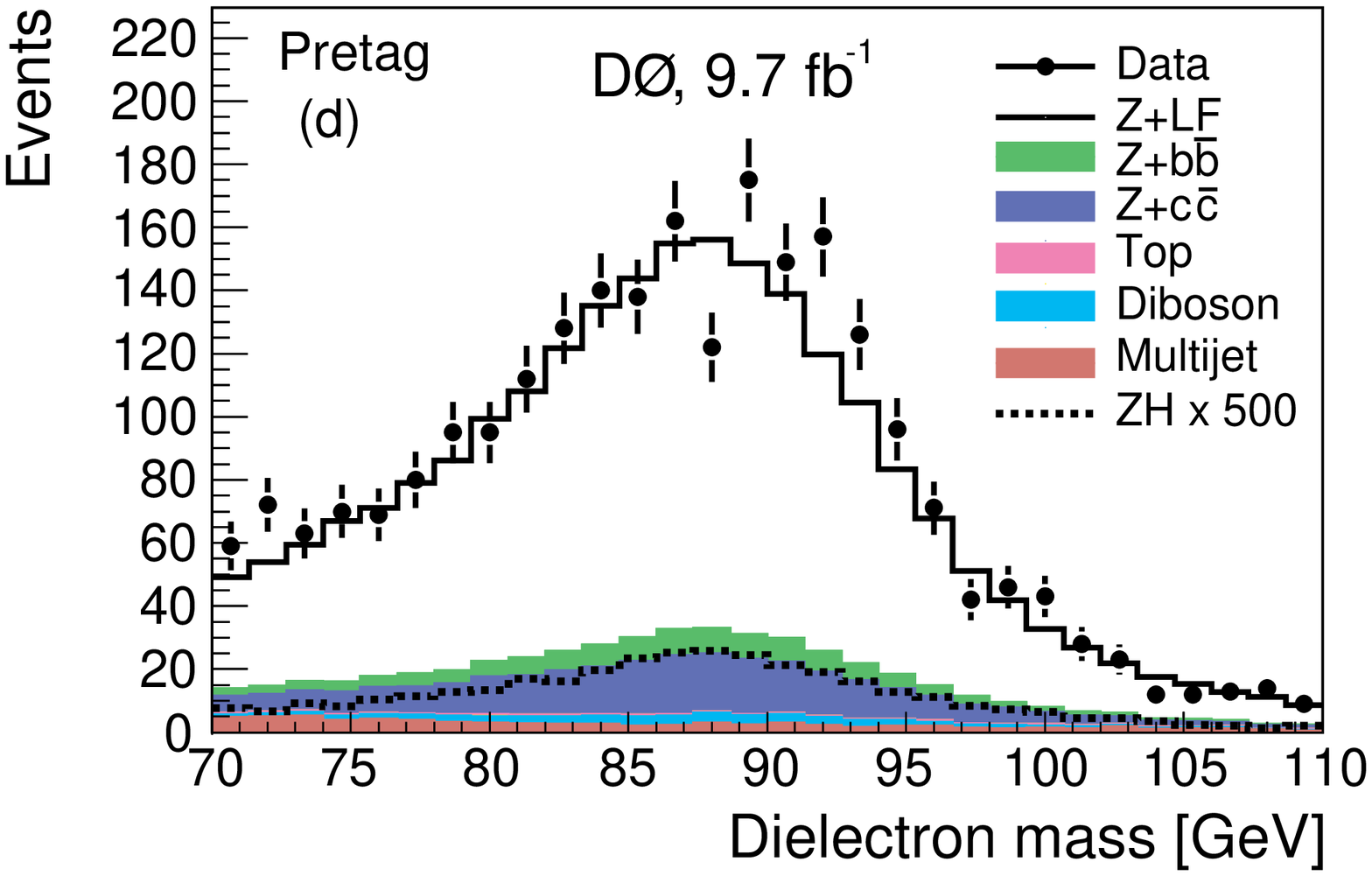} \\
\end{tabular} 
\caption{\label{fig:mll} 
The dilepton mass spectra in the (a) \mumu, (b) \mumutrk, (c) \ee~and 
(d) \eeicr~ channels.
Distributions are shown in the pretag control sample, in which all
selection requirements except $b$-tagging are applied. 
Signal distributions, for $M_H=$ 125 GeV, are scaled by a factor of 500.
}
\end{centering}
\end{figure*}

The dimuon and dielectron mass spectra, after requiring two leptons
and at least two jets are shown in Fig. \ref{fig:mll}.
Distributions of the dijet invariant mass spectra
before and after adjustment by the kinematic fit, are shown in
Fig.~\ref{fig:mbb_posttag}.
A complete list of RF input variables is shown in 
Table~\ref{table:RFvariables}.
Comparisons of the data and MC distributions of the $\ttbar$ RF output
summed over all lepton channels are shown for $M_H=$ 125 GeV in
Figure~\ref{fig:ttbarRF_output}.  
Post-kinematic fit dijet mass distributions for ST and DT in the 
$\ttbar$ depleted region are shown in Fig.~\ref{fig:ttpoor_mbb_fit}.
Fig.~\ref{fig:rf_rich_postfit} displays
the global RF distributions in the $\ttbar$
enriched region, after the fit to the background-only hypothesis.
Fig.~\ref{fig:llr} shows the observed LLR
as a function of Higgs boson
mass.  Also shown are the expected (median) LLRs for the background-only and
signal+background hypotheses, together with the one and two standard
deviation bands about the background-only expectation. 
Fig.~\ref{fig:vz_rf_poor_postfit} shows
the post-fit RF distributions in the  $\ttbar$ depleted region
for the $VZ$ search.  Fig.~\ref{fig:vz_mbb_poor_postfit} displays
the post-fit distribution of the dijet invariant mass from the 
kinematic fit.

\begin{figure*}[tbp]
\begin{centering}
\begin{tabular}{cc}
\includegraphics[height=0.18\textheight]{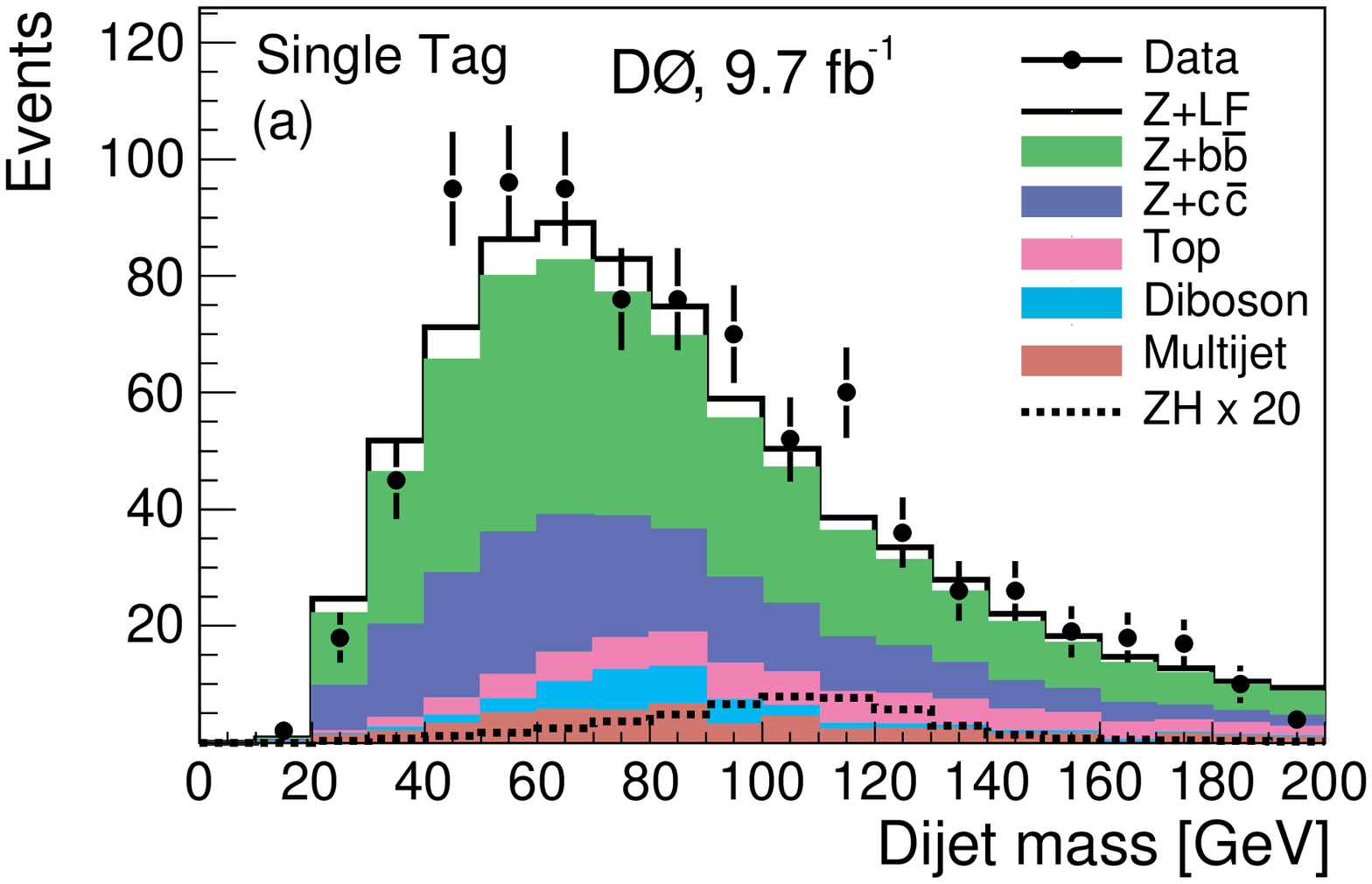}&
\includegraphics[height=0.18\textheight]{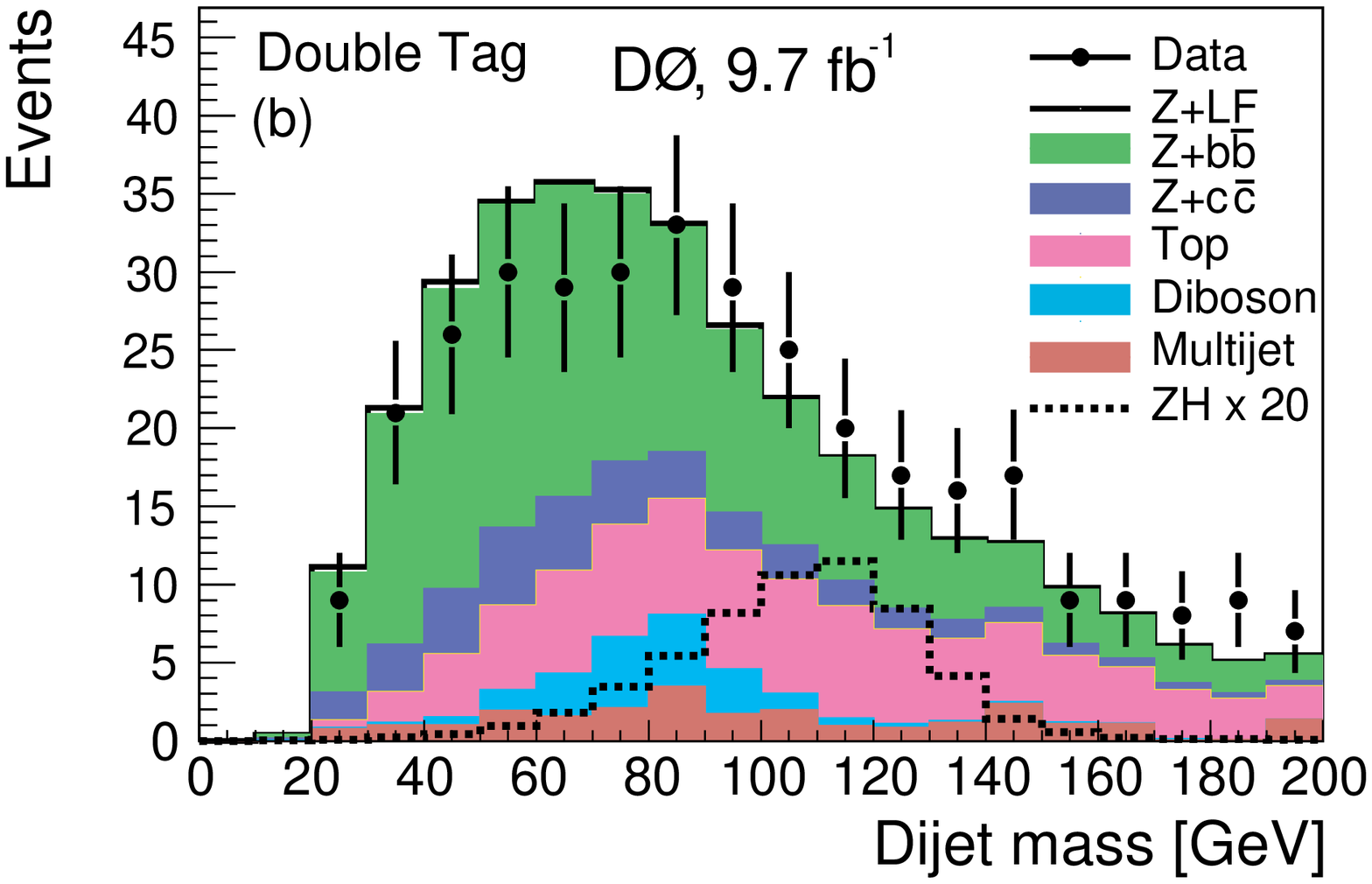}\\
\includegraphics[height=0.18\textheight]{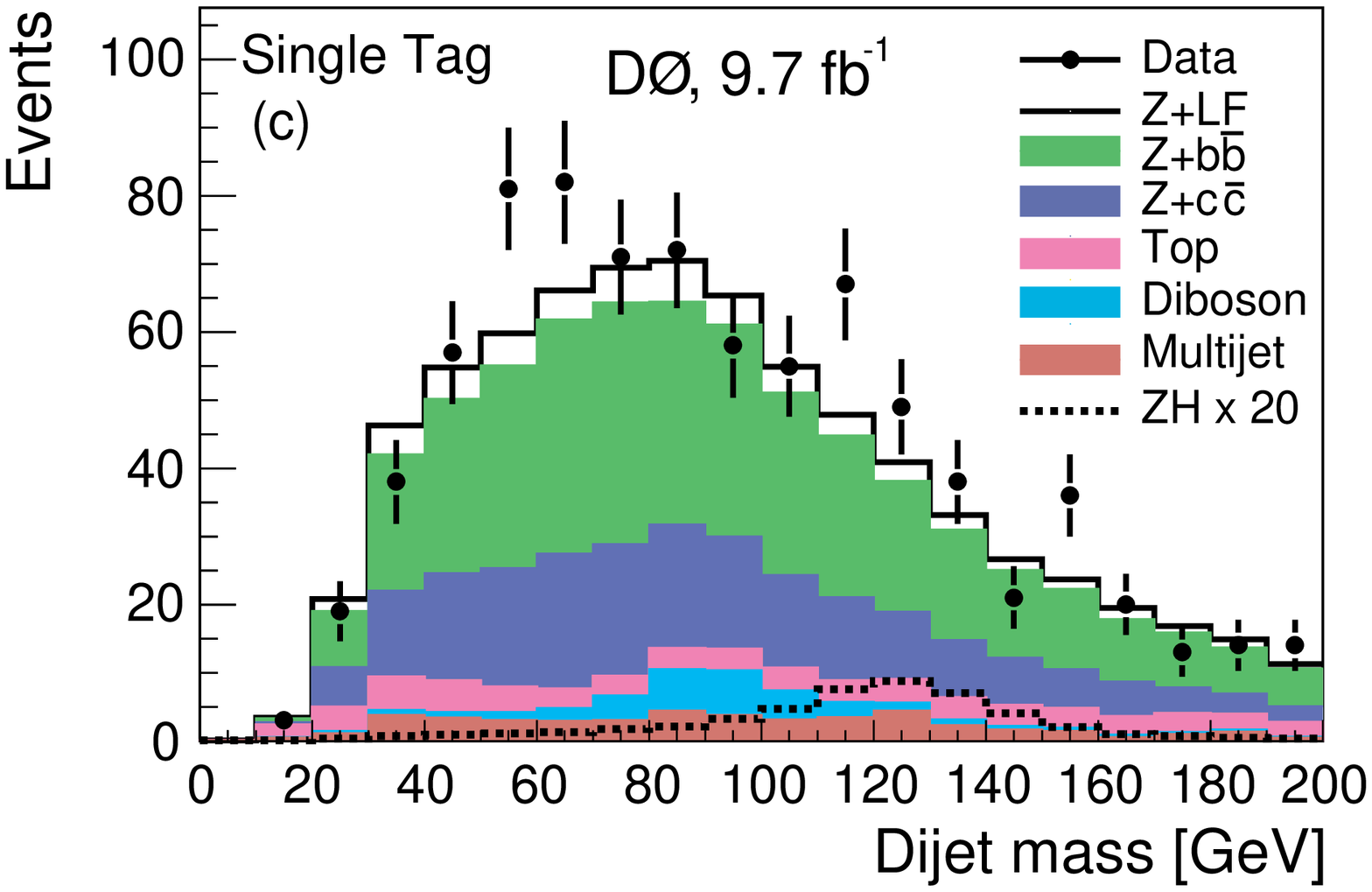}&
\includegraphics[height=0.18\textheight]{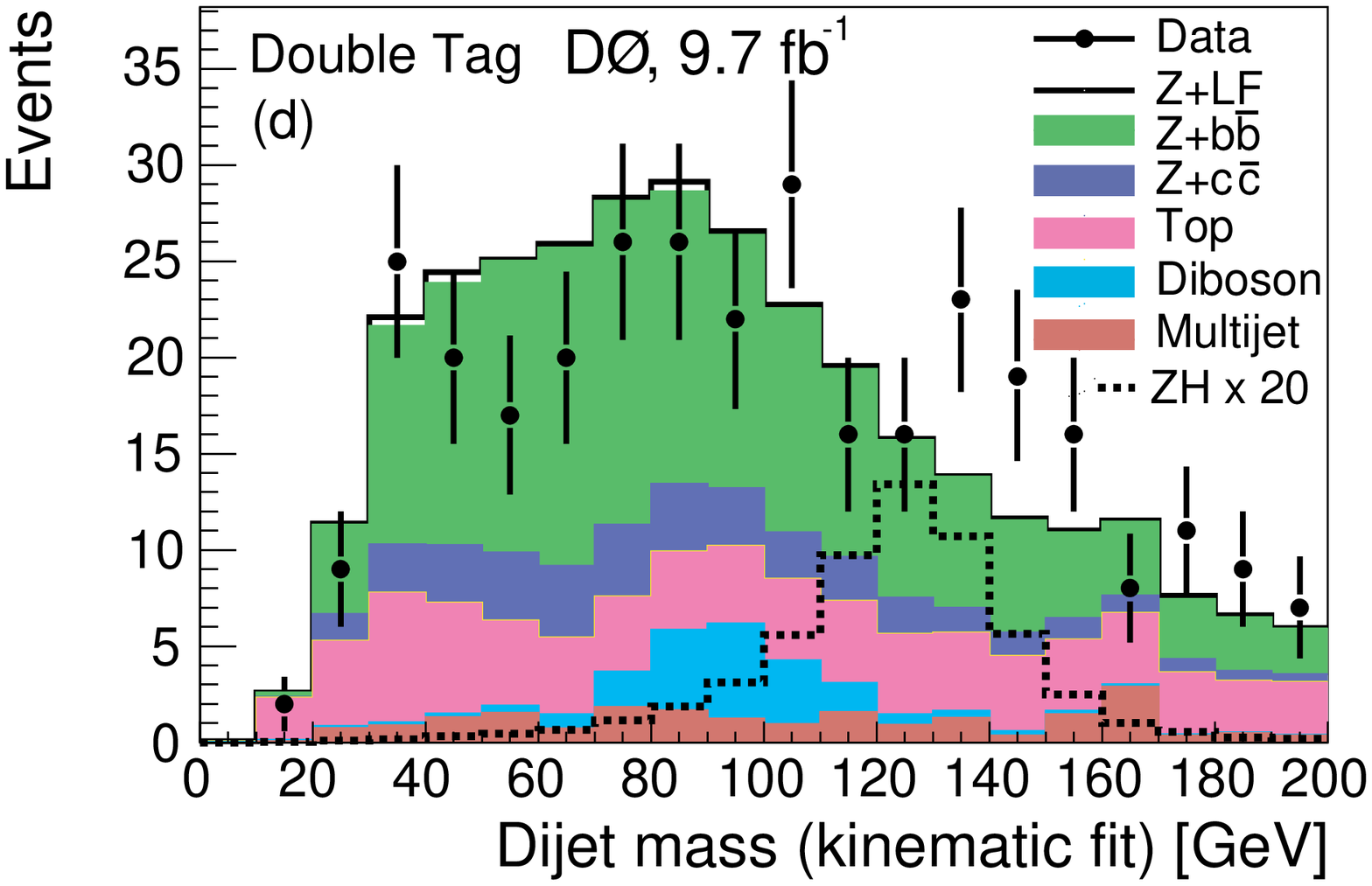}\\
\end{tabular}
\caption{Dijet invariant mass distributions before the kinematic fit in
(a) ST events and (b) DT events; and after the kinematic fit in (c)
ST events and (d) DT events, combined for all lepton channels. 
Signal distributions ($M_{H} = 125$~GeV)
are shown with the SM cross section multiplied by 20.} \label{fig:mbb_posttag}
\end{centering}
\end{figure*}


\begin{table*}[hbt] \centering
\caption{Variables used for the $\ttbar$ and global RF training. The jets 
that form the Higgs boson candidate are referred to as $b_1$ and $b_2$.}
\label{table:RFvariables}
\begin{tabular}{llcc}
\hline
variables&definition&$t\bar{t}$ RF&global RF\\
\hline
{\tt $\mbb$($\mbb^{fit}$)} & invariant mass of the dijet system before (after) the kinematic fit & $\surd$ &$\surd$ \\
{\tt $\ptbone$($\ptbone^{fit}$)} & transverse momentum of the first jet before (after) kinematic fit& $\surd$&$\surd$ \\
{\tt $\ptbtwo$($\ptbtwo^{fit}$)} & transverse momentum of the second jet before (after) kinematic fit &$\surd$ &$\surd$\\
{\tt $\ptbb$}                    & transverse momentum of the dijet system before the kinematic fit & $\surd$ & $\surd$\\
{\tt $\dphibb$}                  & $\Delta\phi$ between the two jets in the dijet system &$-$&$\surd$\\
{\tt $\detabb$}                  & $\Delta\eta$ between the two jets in the dijet system &$-$&$\surd$\\
{\tt $\mjets$} & invariant mass of all jets in the event (the multijet mass)&$\surd$ & $\surd$\\
{\tt $\ptjets$} & transverse momentum of all jets in the event & $\surd$& $\surd$\\
{\tt $\htjets$} & scalar sum of the transverse momenta of all jets in the event &$\surd$ &$-$\\
{\tt $\ptbb/(|\ptbone|+|\ptbtwo|)$} & ratio of dijet system $p_{T}$ over the scalar sum of the $p_{T}$ of the two jets&$\surd$ & $-$\\
{\tt $\mll$} & invariant mass of the dilepton system &$\surd$ &$-$\\
{\tt $\ptz$} & transverse momentum of the dilepton system&$\surd$ &$\surd$ \\
{\tt $\dphill$} & $\Delta\phi$ between the two leptons &$\surd$ &$\surd$\\
{\tt $\colll$} & cosine of the angle between the two leptons (colinearity) &$\surd$ &$\surd$\\
{\tt $\dphizbb$} & $\Delta\phi$ between the dilepton and dijet systems &$\surd$ &$\surd$\\
{\tt $\costhst$} & cosine of the angle between the incoming proton and the $Z$ in the zero momentum frame \footnote{S.~Parke and S.~Veseli,
 Phys. Rev. D {\bf 60}, 093003 (1999).} &$-$&$\surd$\\
{\tt $\mllbb$}   & Invariant mass of dilepton plus dijet system &$-$&$\surd$\\
{\tt $\htllbb$}  & Scalar sum of the transverse momenta of the leptons and jets&$-$&$\surd$\\
{\tt $\met$} & missing transverse energy of the event &$\surd$ &$-$\\
{\tt $\metsig$} &  the $\met$ significance \footnote{A. Schwartzman, FERMILAB-THESIS-2004-21.} &$\surd$ &$\surd$ \\
{\tt $-\ln~L_{fit}$} &  negative log likelihood from the kinematic fit&$\surd$ &$\surd$\\
$\ttbar$ RF                & $\ttbar$ RF output & $-$&$\surd$\\
\hline
\end{tabular} 
\end{table*}


\begin{figure*}[htdp]\centering
\begin{tabular}{cc}
\includegraphics[height=0.2\textheight]{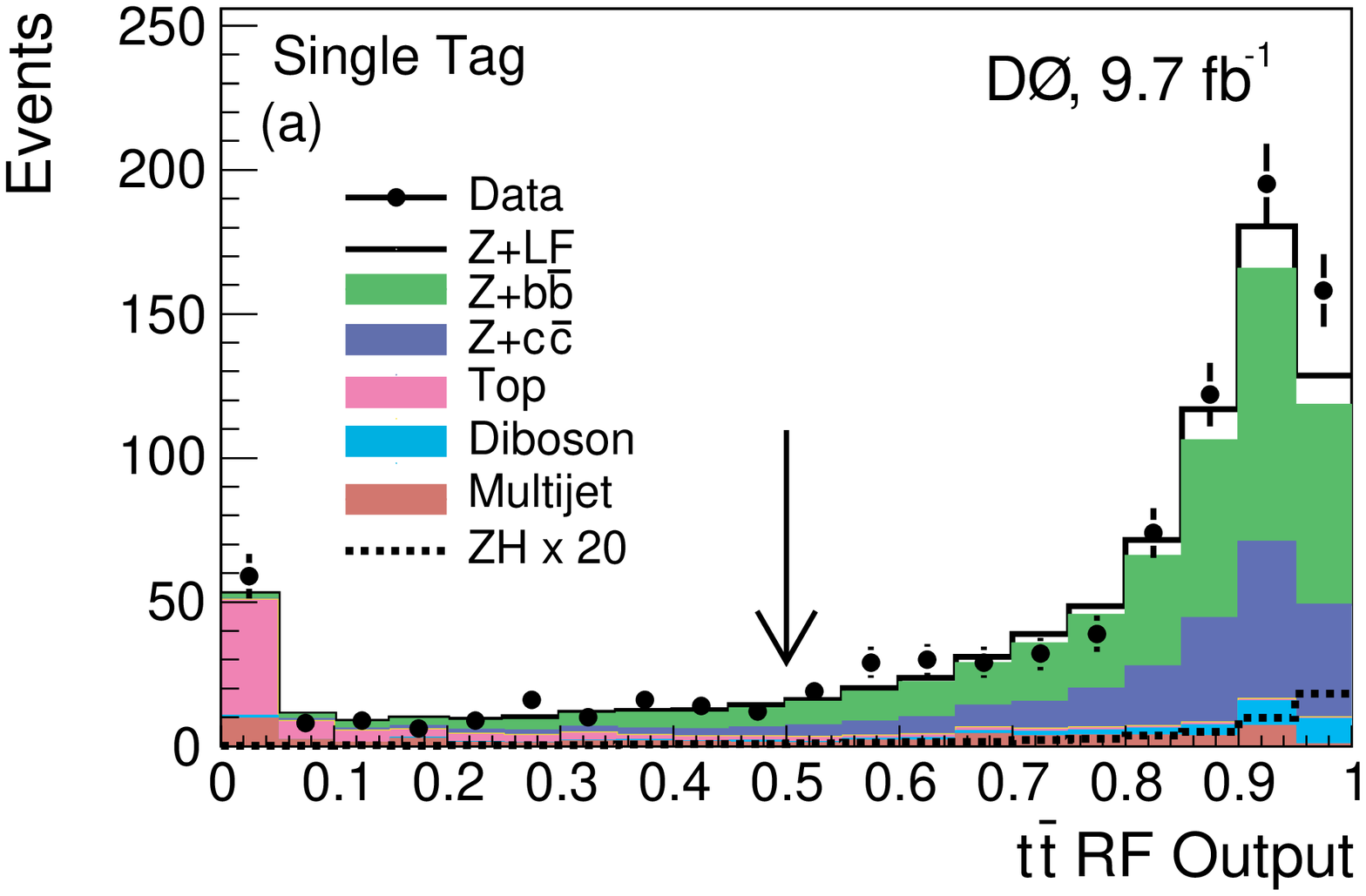} &
\includegraphics[height=0.2\textheight]{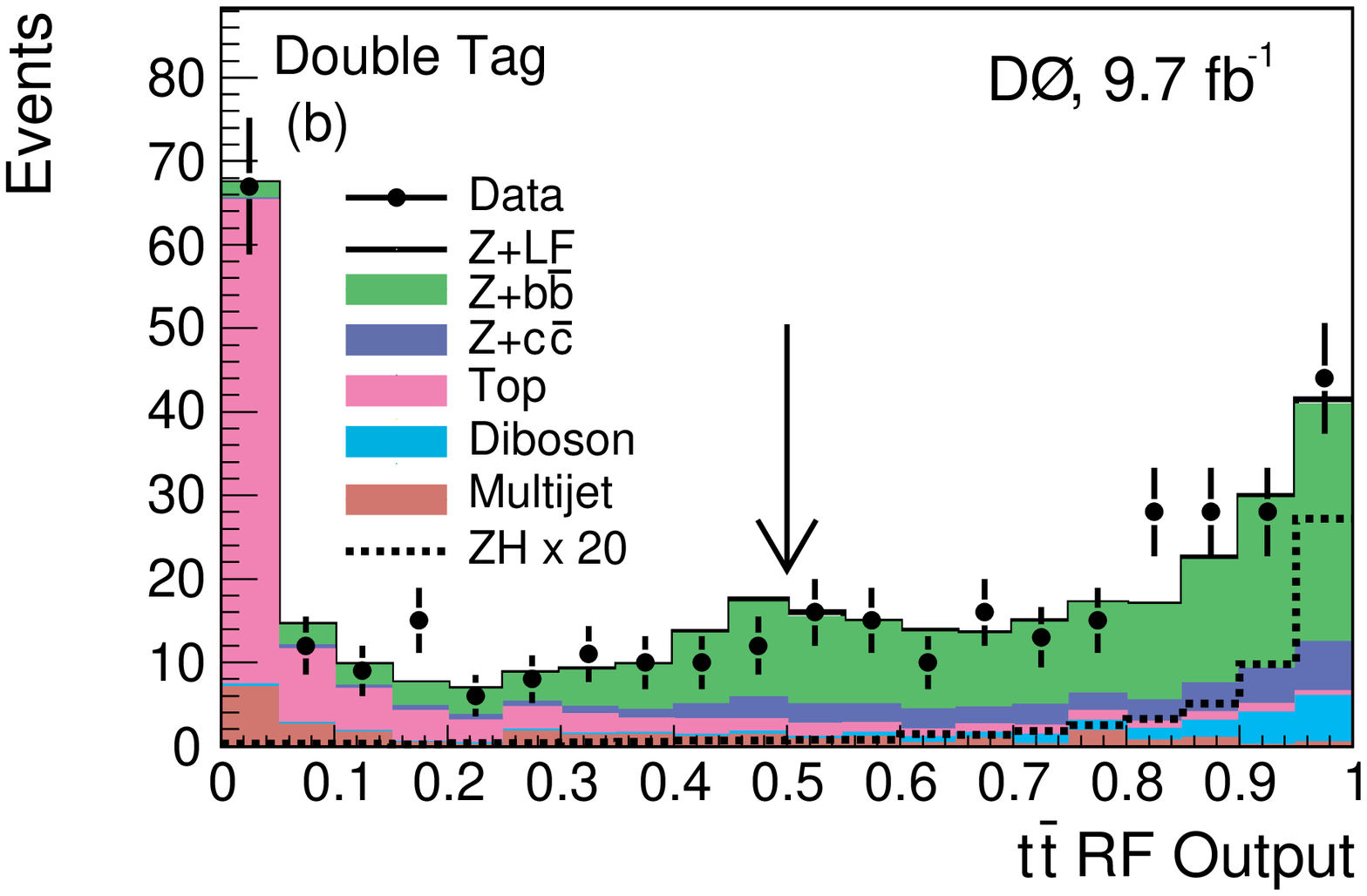} \\
\end{tabular}
\caption[$\ttbar$ RF output for all channels combined and $M_H=125~\GeV$]{The
$\ttbar$ RF output ($M_H=125~\GeV$) for all lepton channels combined
(a) ST and (b) DT events. Signal distributions
are shown with the SM cross section multiplied by 20. The vertical arrows
indicate the $\ttbar$~RF selection requirement used to define the $\ttbar$~ 
enriched and depleted samples.
\label{fig:ttbarRF_output}}
\end{figure*}


\begin{figure*}[htbp]
\centering
\begin{tabular}{cc}
\includegraphics[height=0.21\textheight]{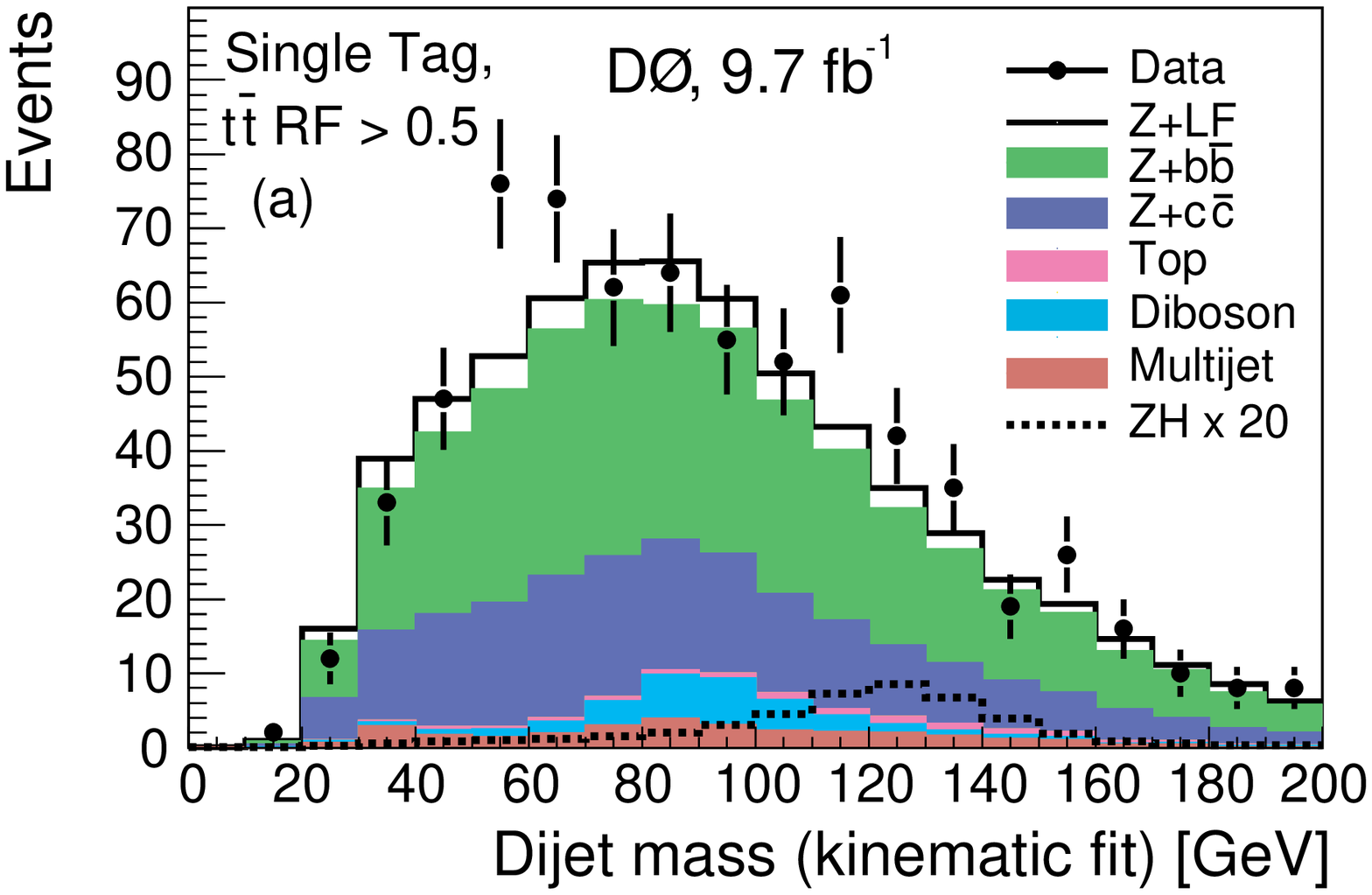}  &
\includegraphics[height=0.21\textheight]{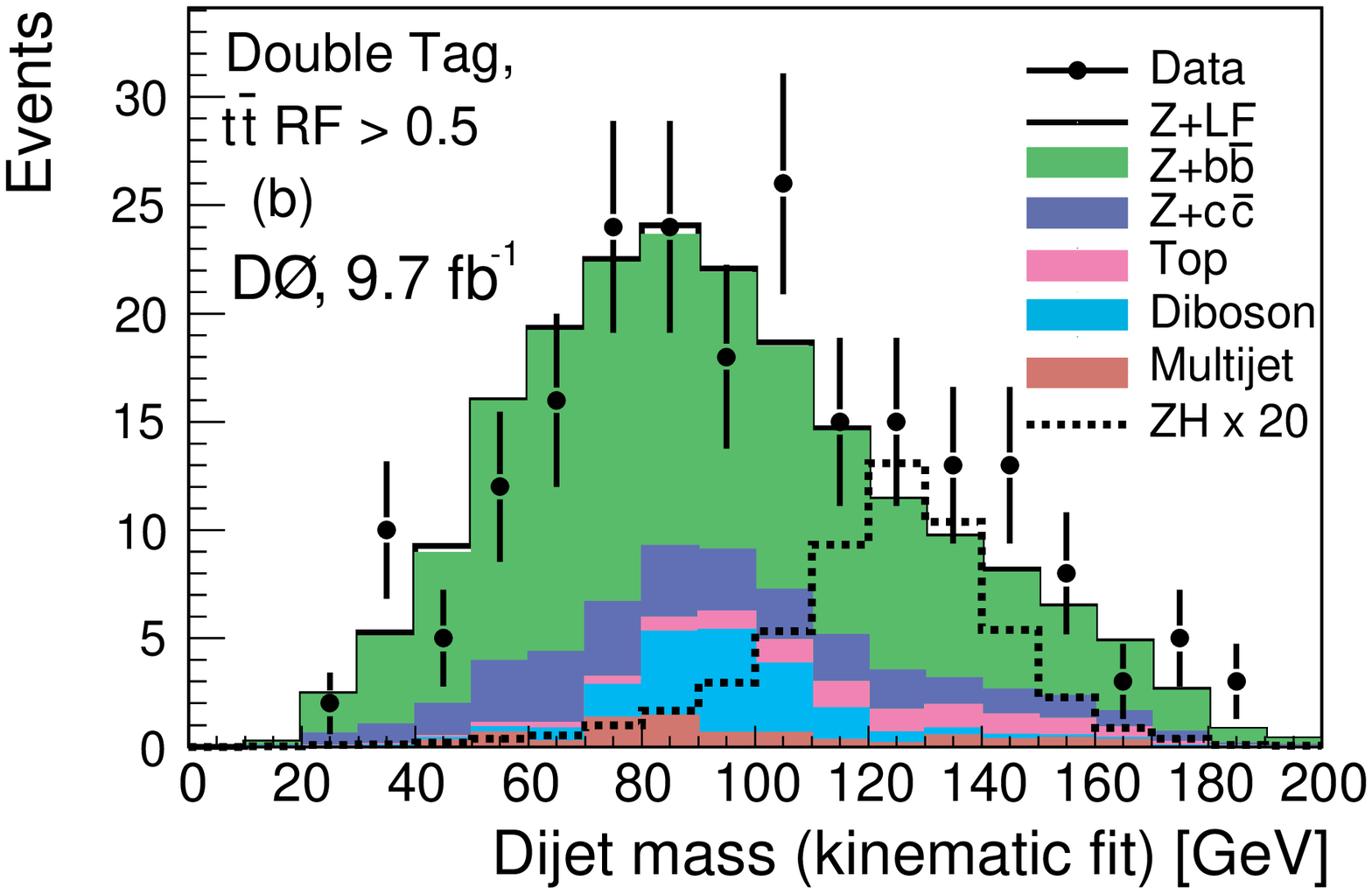} \\
\end{tabular} 
\caption{\label{fig:ttpoor_mbb_fit} Post-kinematic fit dijet mass
distributions
  in the $t\bar{t}$ depleted region for all lepton
  channels combined assuming $M_H=125$ GeV
  for (a) ST events and (b) DT events.
  Signal distributions are shown with the SM cross section multiplied
  by 20.} 
\end{figure*}


\begin{figure*}[htbp]
\centering
\begin{tabular}{cc}
\includegraphics[height=0.21\textheight]{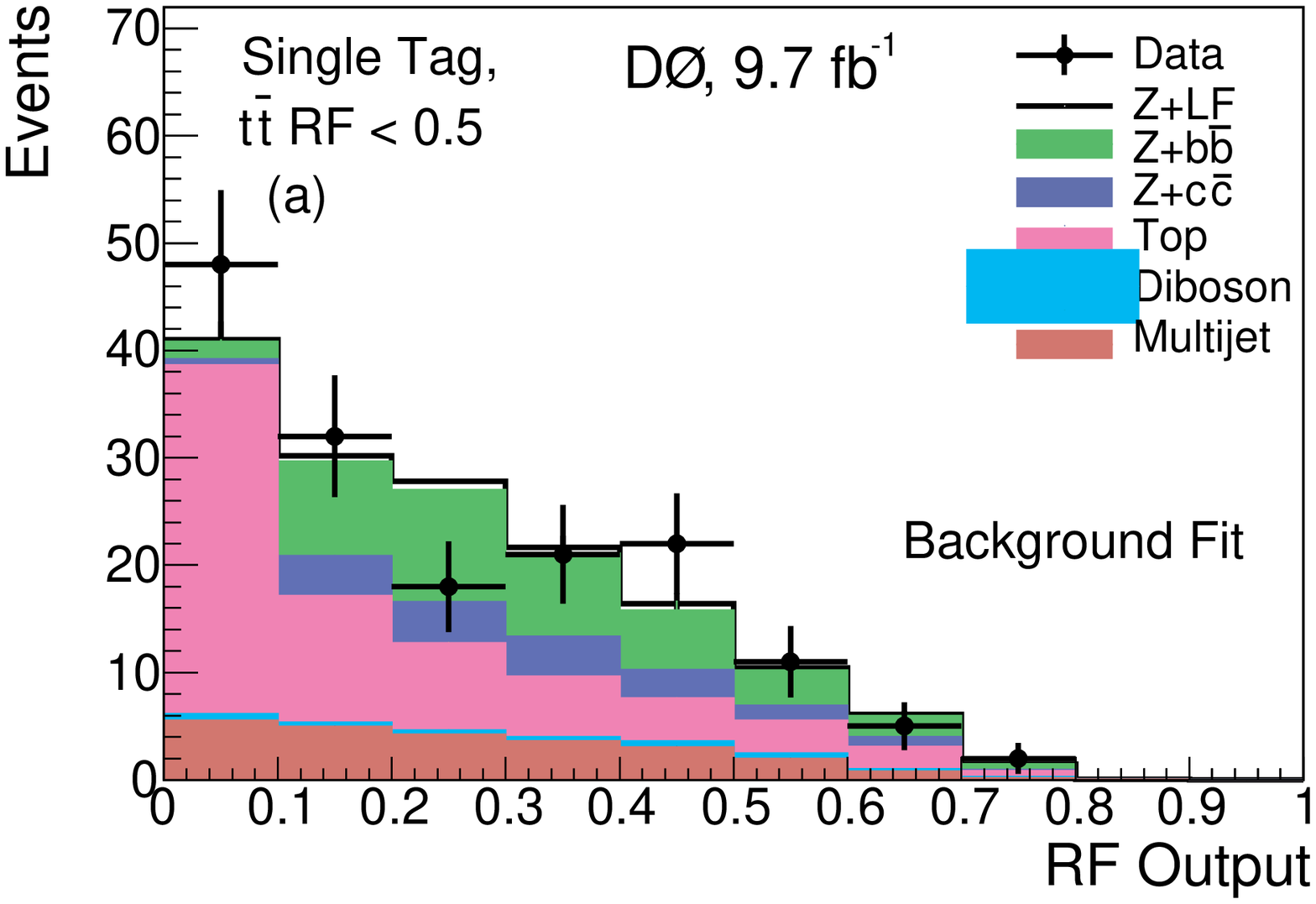}  &
\includegraphics[height=0.21\textheight]{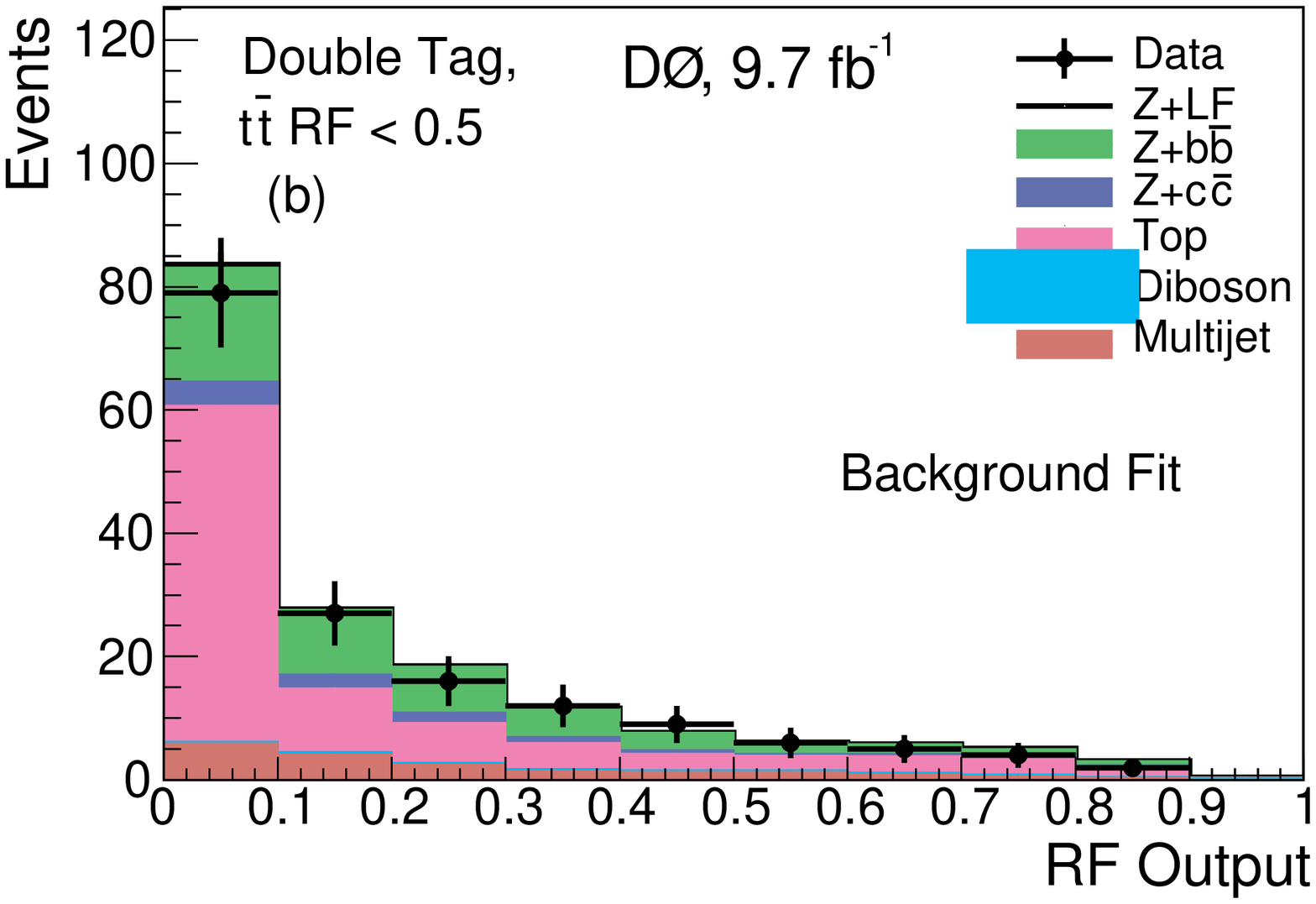} \\
 \includegraphics[height=0.21\textheight]{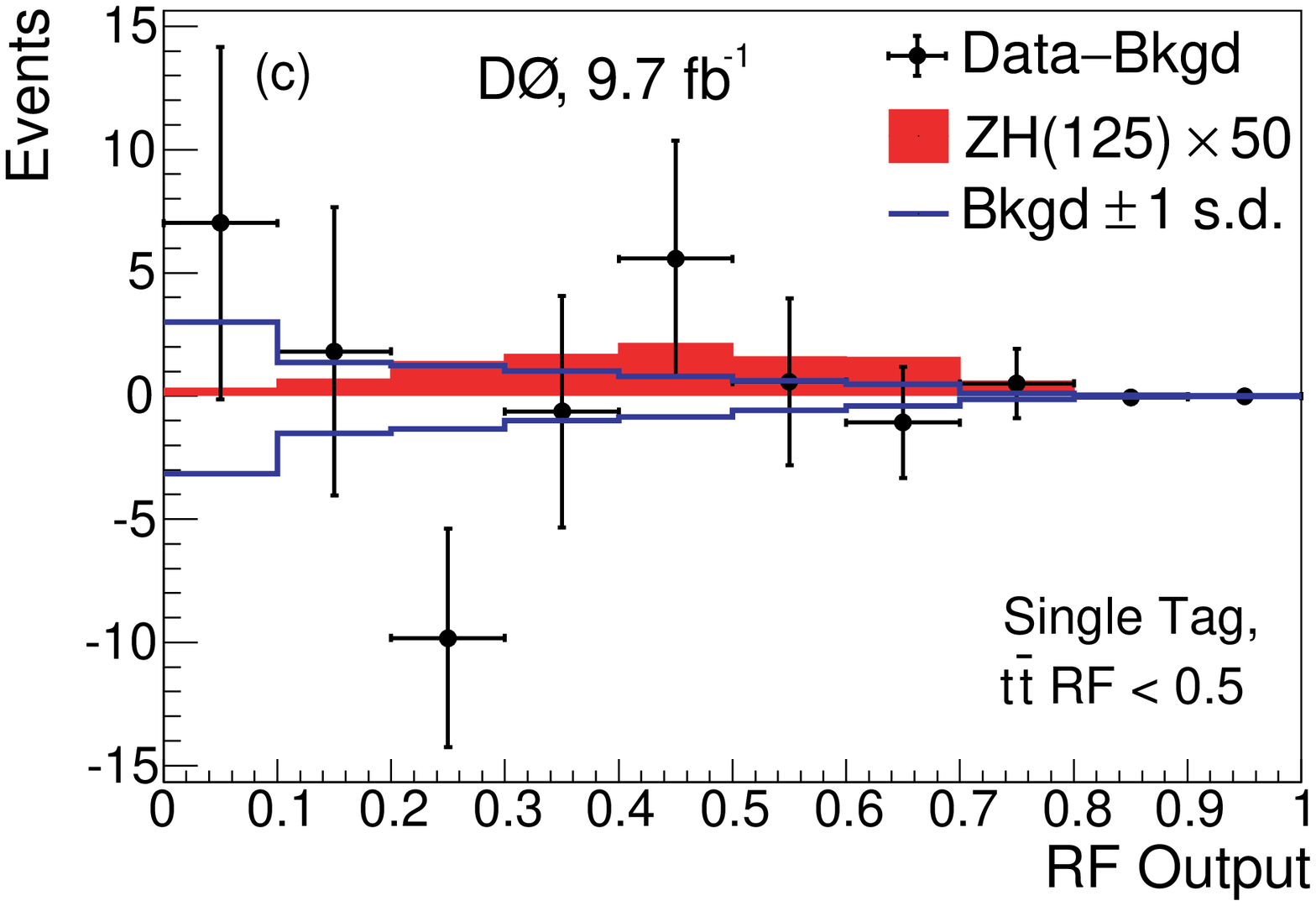}  &
 \includegraphics[height=0.21\textheight]{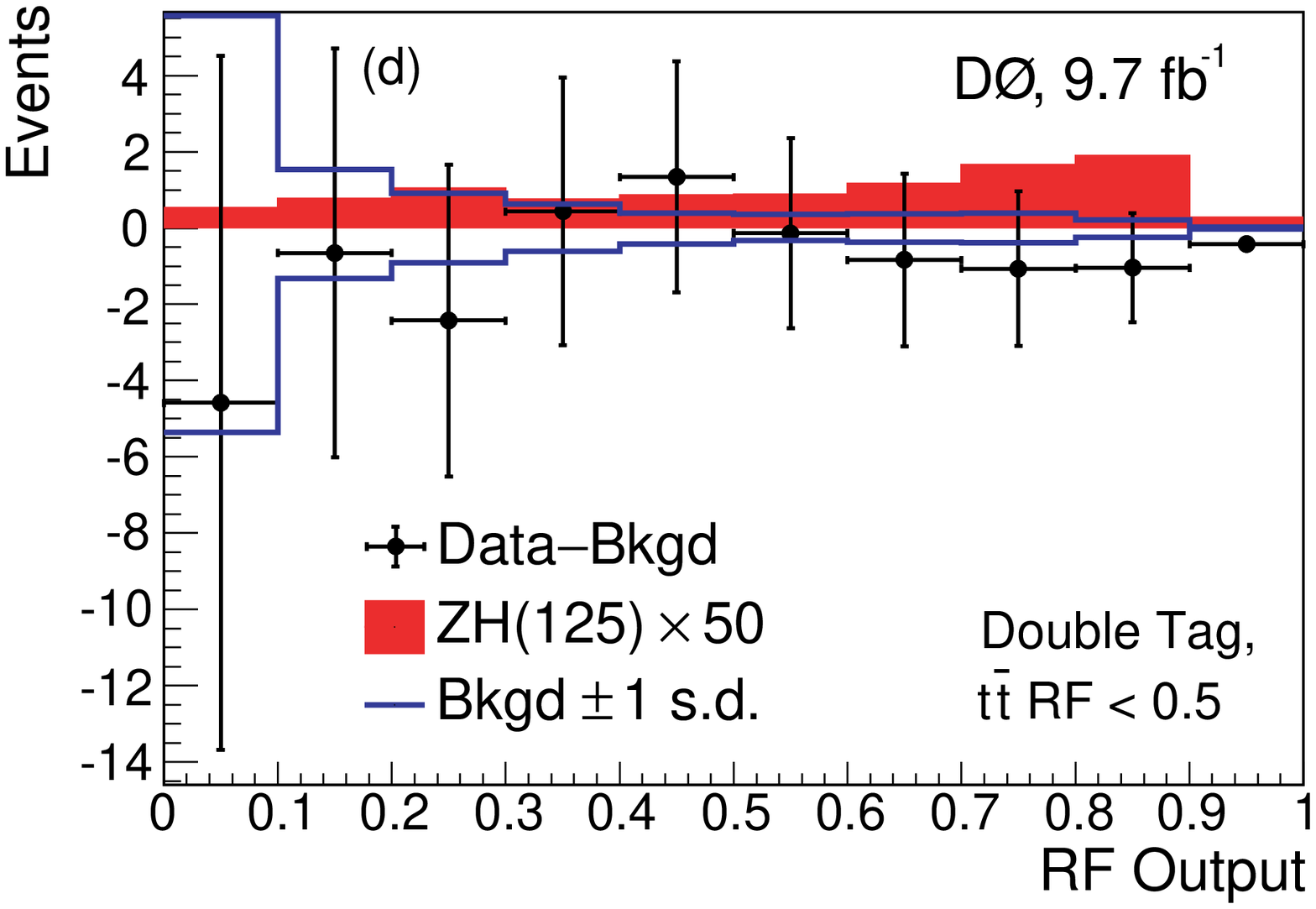} \\
\end{tabular} 
\caption{\label{fig:rf_rich_postfit} Post-fit RF output distributions in the $\ttbar$
  enriched region, assuming $M_H=$ 125 GeV,
  after the fit to the background-only model for (a) ST events
  and (b) DT events. Background-subtracted distributions for (a) and (b) are shown in 
  (c) and (d), respectively. 
  Signal distributions are shown with the SM cross section 
  scaled to 50 $\times$ SM prediction in (c) and (d).   
  The blue lines are the total posterior systematic uncertainty following a fit of 
  the background to the data.}
\end{figure*}


\begin{figure*}[tbp]
\centering
\includegraphics[height=0.25\textheight]{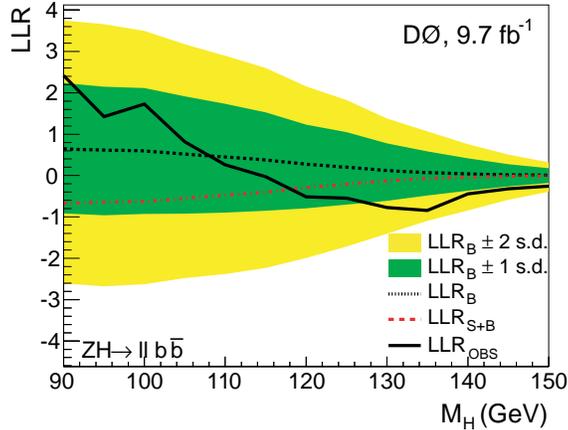}
\caption{Observed LLR as a function of Higgs boson mass.  Also shown are 
the expected LLRs for the background-only (B) and signal+background
(S+B) hypotheses, together with the one and two standard deviation (s.d.)
bands about the background-only expectation.  
}
\label{fig:llr}
\end{figure*}


\begin{figure*}[htbp]
\centering
\begin{tabular}{cc}
\includegraphics[height=0.21\textheight]{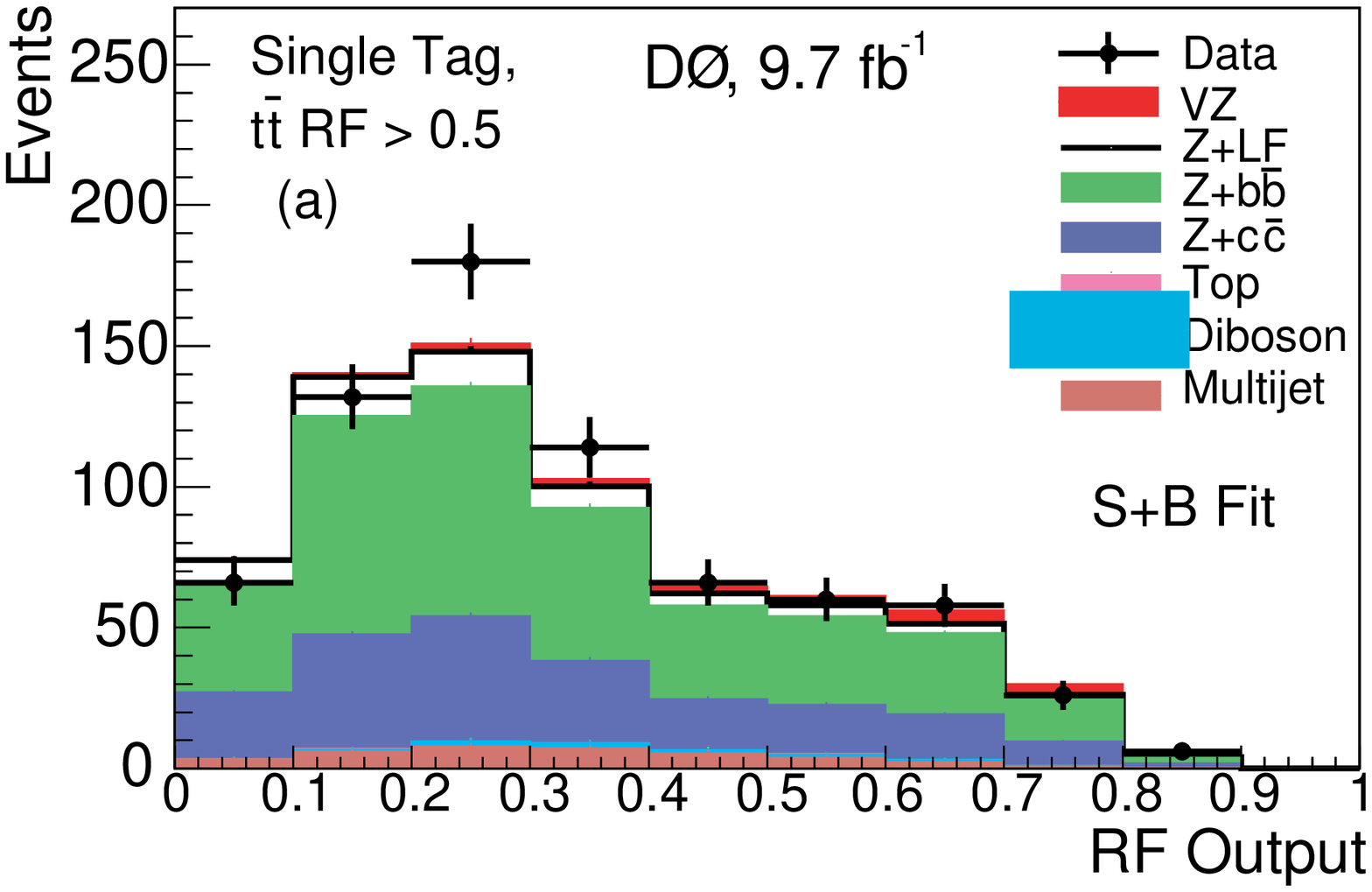}  &
\includegraphics[height=0.21\textheight]{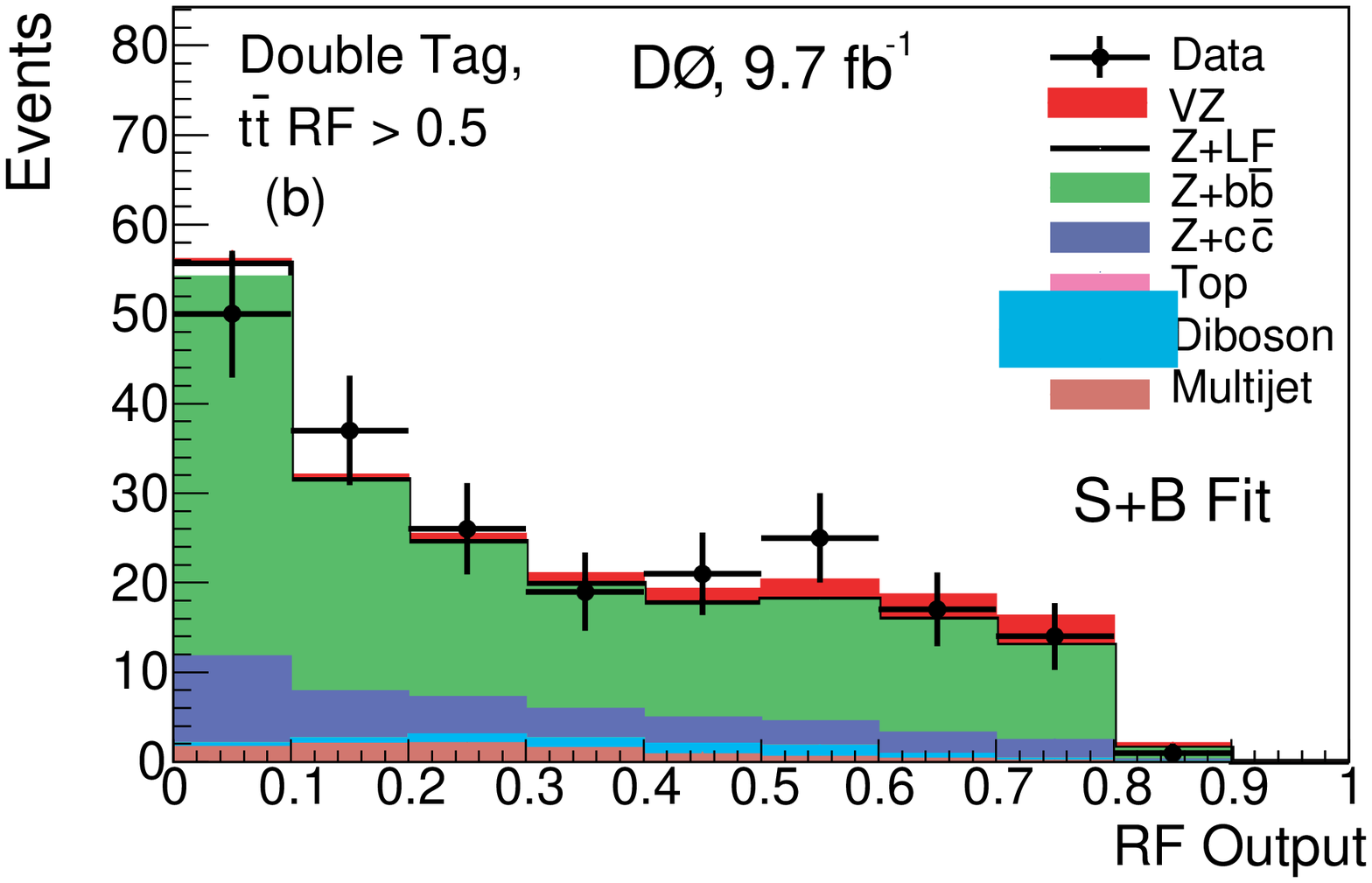} \\
 \includegraphics[height=0.21\textheight]{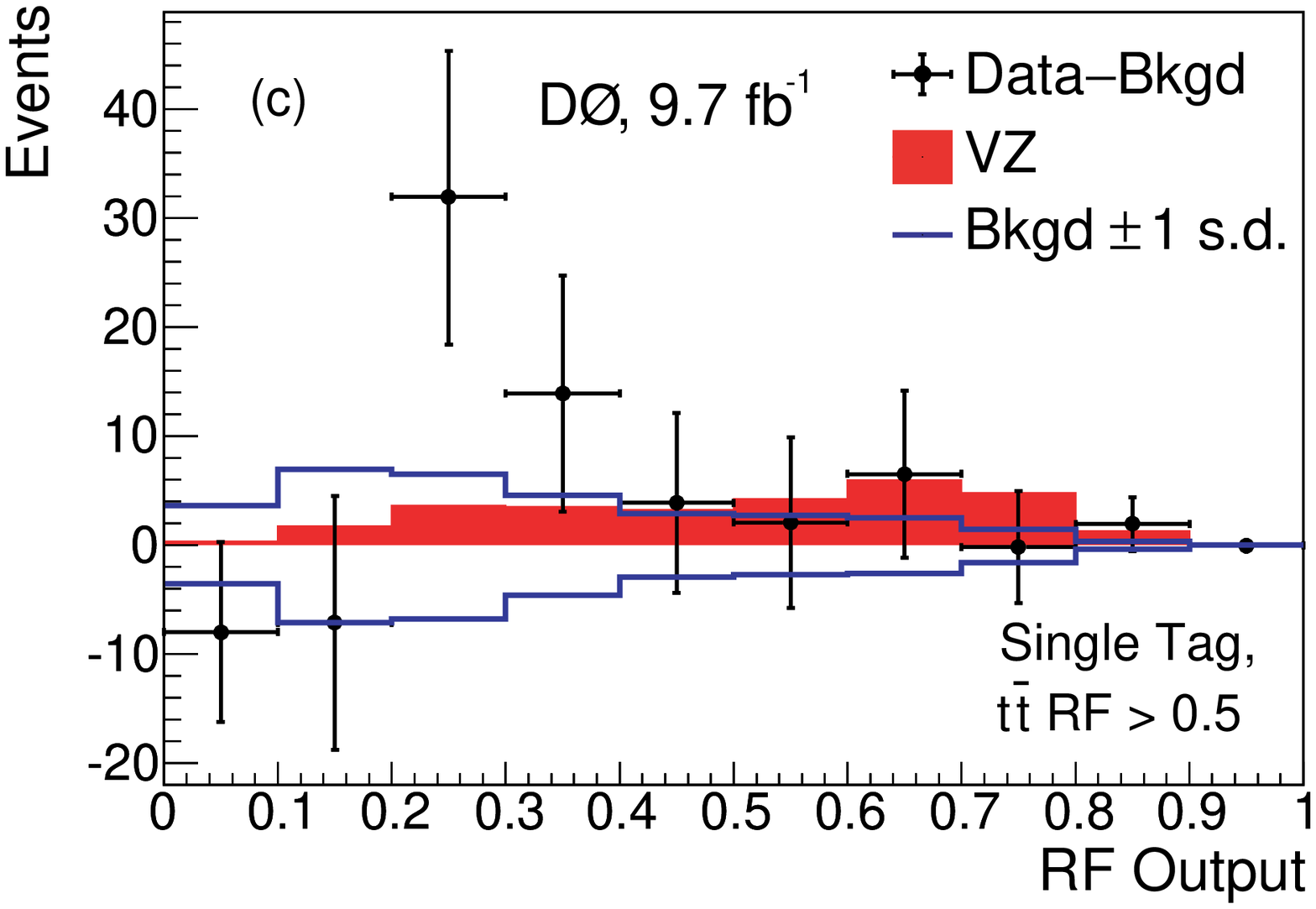}  &
 \includegraphics[height=0.21\textheight]{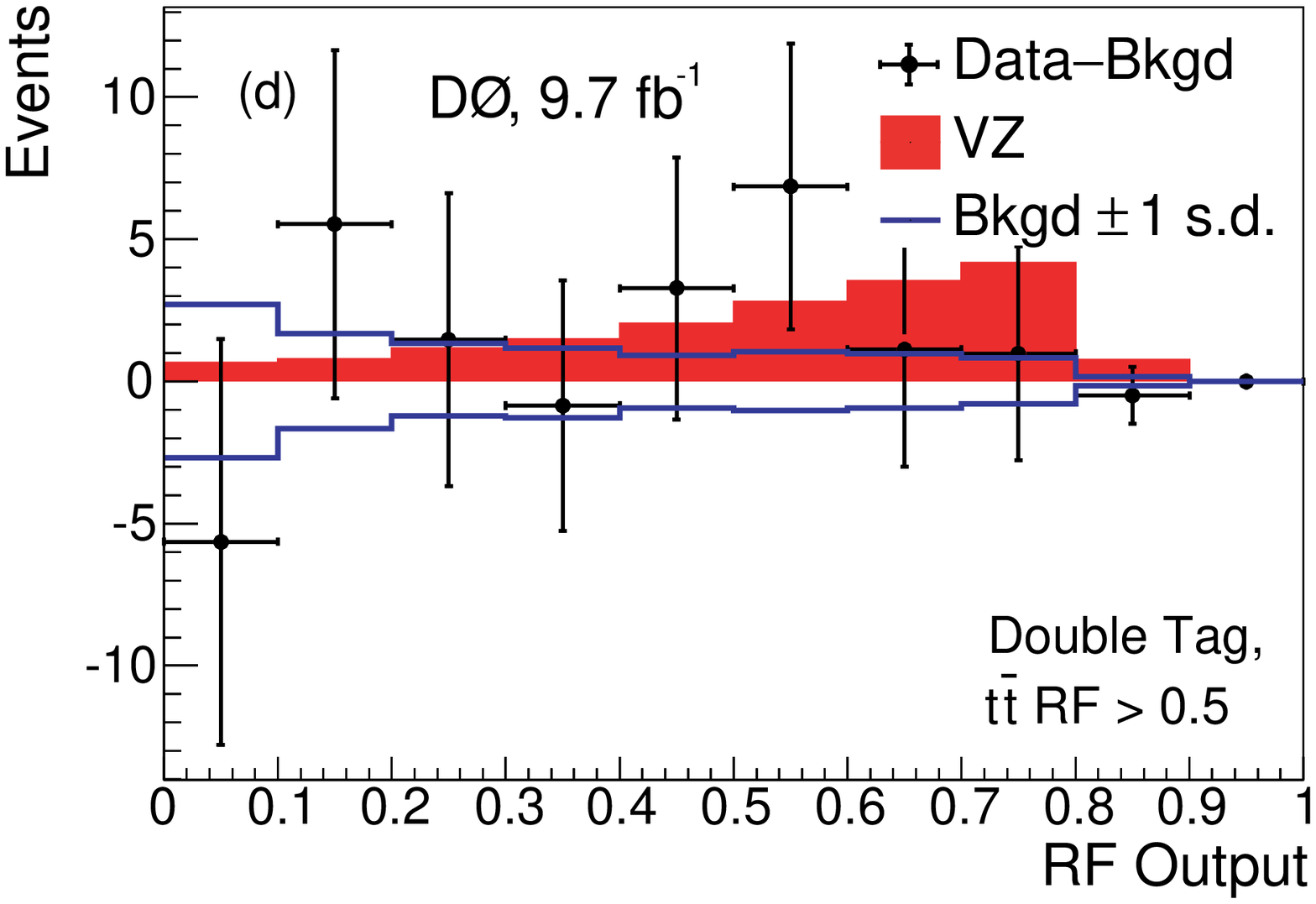} \\
\end{tabular} 
\caption{\label{fig:vz_rf_poor_postfit} Post-fit $VZ$ RF output
  distributions in the $\ttbar$ depleted region after the fit to the
  S+B model for (a) ST events and (b) DT events. Background-subtracted
  distributions for (a) and (b) are shown in (c) and (d),
  respectively.  Signal distributions are scaled to the measured
  $VZ$ cross section.  The blue lines indicate the uncertainty from
  the fit.}
\end{figure*}

\begin{figure*}[htbp]
\centering
\begin{tabular}{cc}
\includegraphics[height=0.21\textheight]{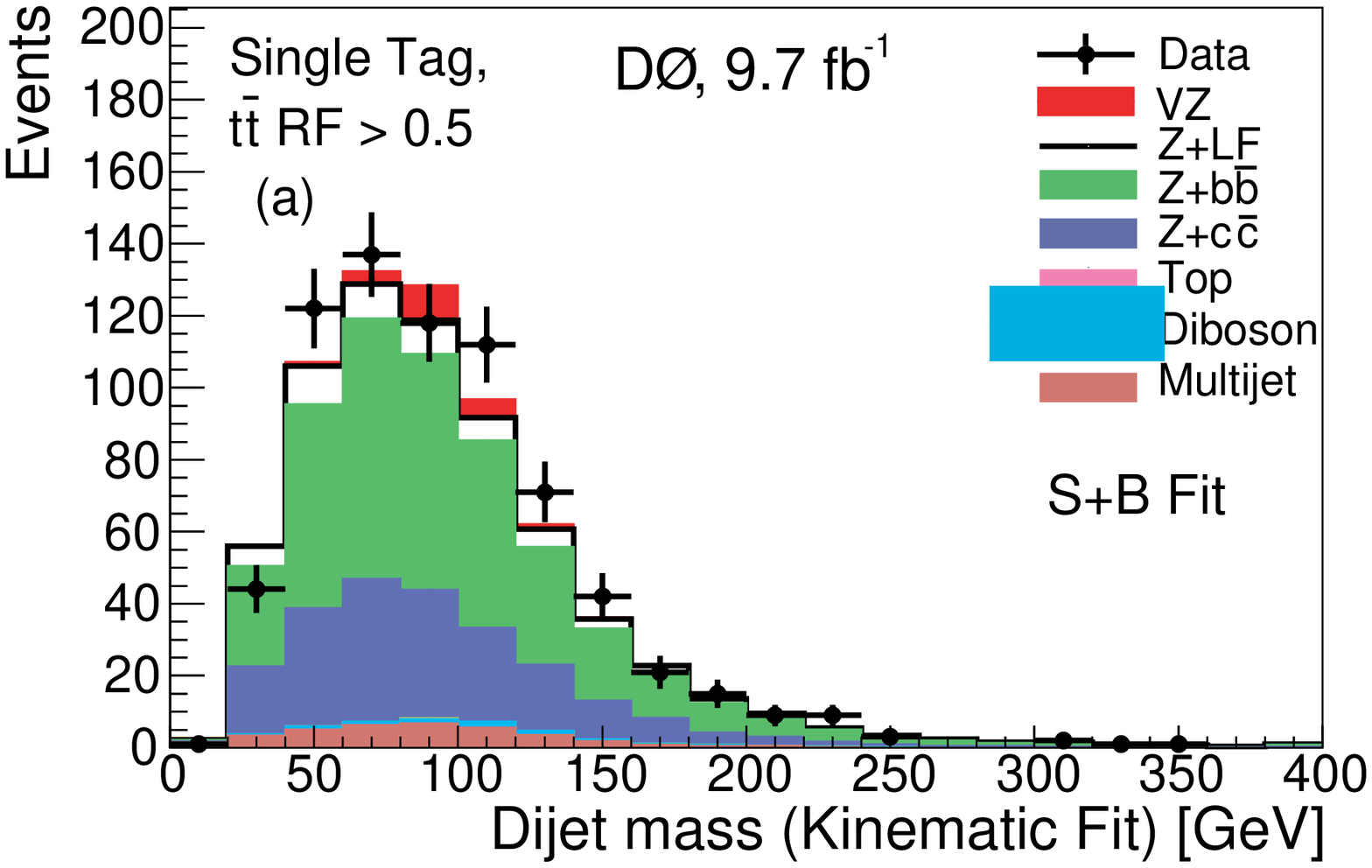}  &
\includegraphics[height=0.21\textheight]{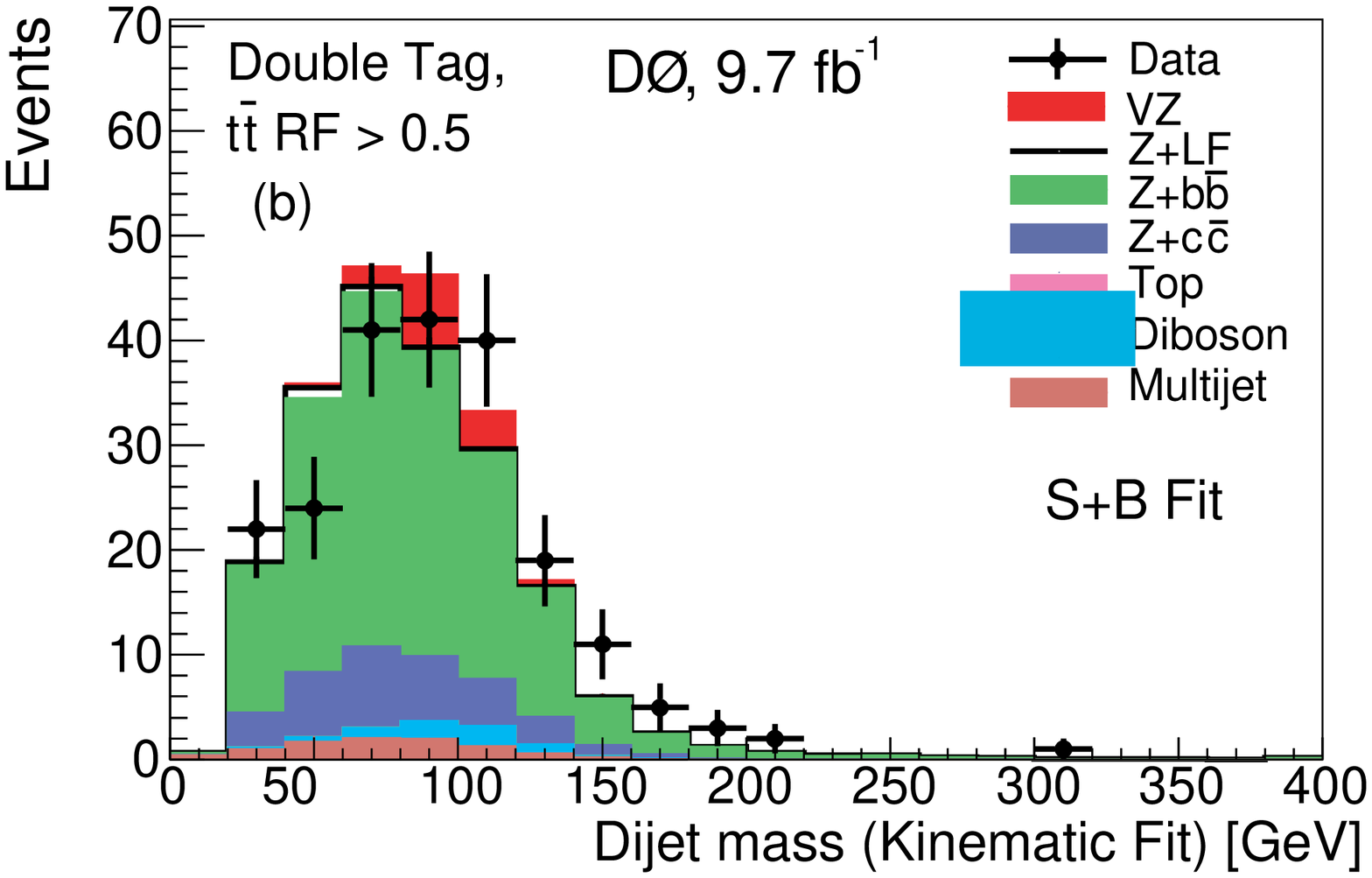} \\
 \includegraphics[height=0.21\textheight]{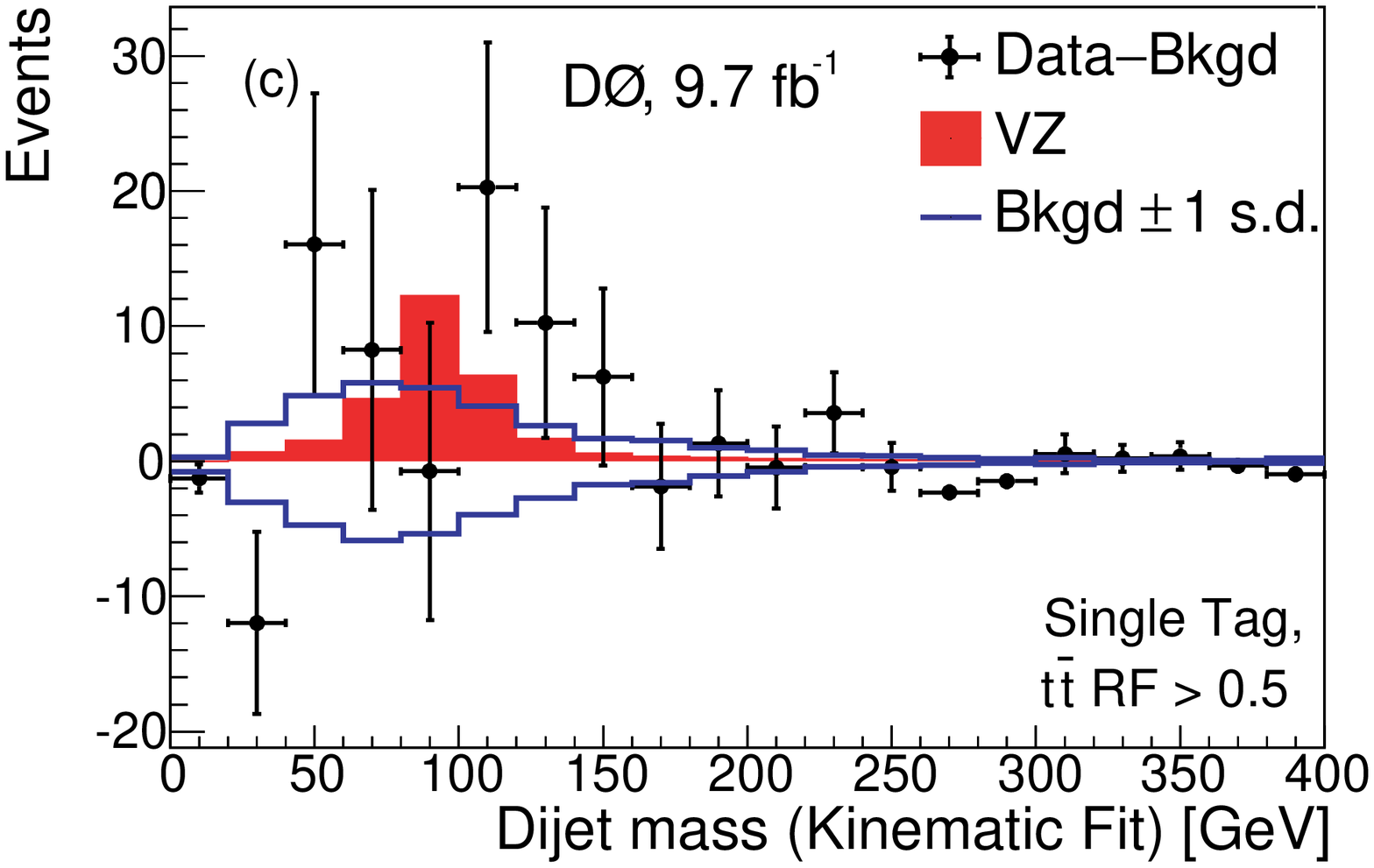}  &
 \includegraphics[height=0.21\textheight]{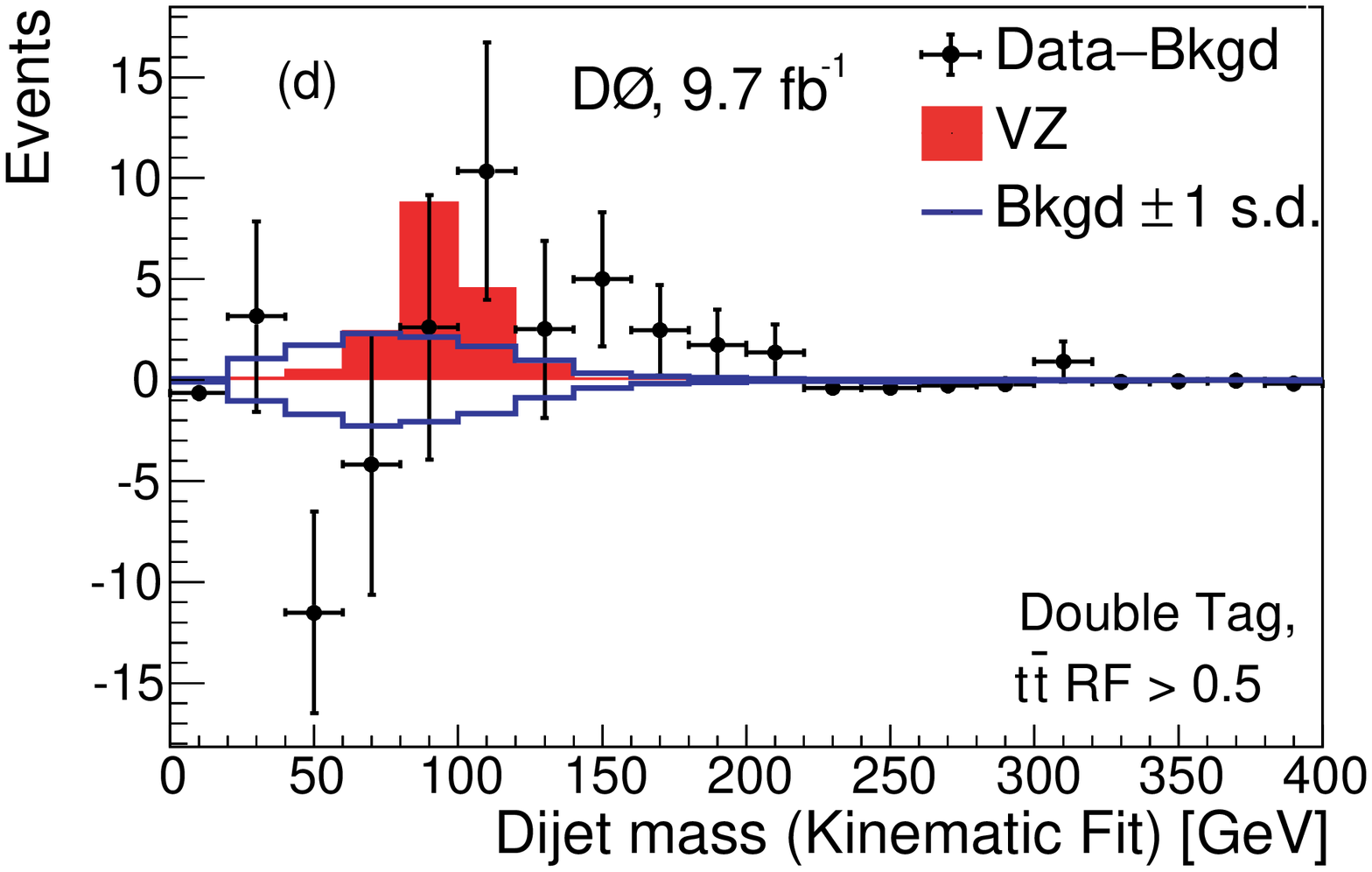} \\
\end{tabular} 
\caption{\label{fig:vz_mbb_poor_postfit} Post-fit distributions of the
  dijet invariant mass (from the kinematic fit) in the $\ttbar$
  depleted region after the fit to the S+B model for (a) ST events and
  (b) DT events. Background-subtracted distributions for (a) and (b)
  are shown in (c) and (d), respectively.  Signal distributions are
  scaled to the measured $VZ$ cross section.  The blue lines
  indicate the uncertainty from the fit.}
\end{figure*}

\end{widetext}



%% file: zh_llbb_prl.bbl
\begin{thebibliography}{99}

\bibitem{higgs} F.~Englert and R.~Brout, Phys. Rev. Lett. {\bf 13}, 321 (1964);
P.W.~Higgs, Phys. Rev. Lett. {\bf 13}, 508 (1964);
G.S.~Guralnik, C.R.~Hagen, and T.W.B.~Kibble, Phys. Rev. Lett. {\bf 13},
585 (1964). 

\bibitem{cdf-wmass} T.~Aaltonen {\sl et al.} (CDF Collaboration), Phys. Rev. Lett. {\bf 108}, 151803 (2012).

\bibitem{d0-wmass} V.M.~Abazov {\sl et al.} (D0 Collaboration), Phys. Rev. Lett. {\bf 108}, 151804 (2012).

\bibitem{tev-top} 
  T.~Aaltonen {\sl et al.}  [CDF and D0 Collaborations],
  arXiv:1207.1069 [hep-ex].

\bibitem{ew-fit}
LEP Electroweak Working Group \\
http://lepewwg.web.cern.ch/LEPEWWG/.

\bibitem{lep-results}
ALEPH, DELPHI, L3, and OPAL Collaborations,
Phys. Lett. B {\bf 565}, 61 (2003).

\bibitem{tev-results}
Tevatron New Phenomena and Higgs Working Group,\\
arXiv:1203.3774.

\bibitem{atlas-results} 
G.~Aad {\sl et al.} (ATLAS Collaboration), Phys. Rev. D {\bf 86}, 032003 (2012).

\bibitem{cms-results}
S.~Chatrchyan {\sl et al.} (CMS Collaboration), Phys. Lett. B {\bf 710}, 26 (2012).

\bibitem{new-LHC-results}
G.~Aad {\it et. al.} (ATLAS Collaboration),
Phys. Lett. B {\bf 716}, 1 (2012);
S.~Chatrchyan {\it et. al.} (CMS Collaboration),
Phys. Lett. B {\bf 716}, 30 (2012).

\bibitem{hbr} 
A.~Djouadi, J.~Kalinowski, and M.~Spira, Comput. Phys. Commun. {\bf 108}, 56 (1998);
A.~Bredenstein, A.~Denner, S.~Dittmaier, and M.M.~Weber, Phys. Rev. D {\bf 74} 013004 (2006).

\bibitem{d0-zhiggs}
V.M.~Abazov {\sl et al.} (D0 Collaboration), Phys. Rev. Lett. {\bf 105}, 251801 (2010).

\bibitem{cdf-zhiggs}
T.~Aaltonen {\sl et al.} (CDF Collaboration), Phys. Rev. Lett. {\bf 105}, 251802 (2010); Phys. Rev. Lett. {\bf 109}, 111803 (2012).

\bibitem{d0det}
V.M.~Abazov {\sl et al.} (D0 Collaboration), Nucl. Instrum. Methods Phys. Res. A {\bf 565}, 463 (2006);
S.~Abachi {\sl et al.} (D0 Collaboration), Nucl. Instrum. Methods Phys. Res. A {\bf 338}, 185 (1994).

\bibitem{d0upgrade}
M.~Abolins {\sl et al.}, Nucl. Instrum. Methods Phys. Res. A {\bf 584}, 75 (2008); 
R.~Angstadt {\sl et al.}, Nucl. Instrum. Methods Phys. Res. A {\bf 622}, 298 (2010);
S.N.~Ahmed {\sl et al.}, Nucl. Instrum. Methods Phys. Res. A {\bf 634}, 8 (2011).

\bibitem{runiicone} 
G. C. Blazey {\sl et al.}, arXiv:hep-ex/0005012.

\bibitem{appendix}
See Appendix.

\bibitem{bid}
V.M.~Abazov {\sl et al.} (D0 Collaboration), 
Nucl. Instrum. Methods Phys. Res. A {\bf 620}, 490 (2010).
 
\bibitem{pythia}
T.~Sj\"{o}strand, S.~Mrenna, and P.~Skands, J. High Energy Phys. {\bf 05} (2006) 026; we use version 6.409, D0 Tune A.

\bibitem{alpgen}
M.L.~Mangano, F.~Piccinini, A.D.~Polosa, M.~Moretti, and R.~Pittau, 
J. High Energy Phys. {\bf 07} (2003) 001.
We use version 2.11.

\bibitem{mlm}
S.~H\"{o}che {\sl et al.},
arXiv:hep-ph/0602031.

\bibitem{cteq6}
J.~Pumplin, D.R.`Stump, J.~Huston, H.-L.~Lai, P.~Nadolsky, and W.-K.~Tung, 
J. High Energy Phys. {\bf 07} (2002) 012.

\bibitem{geant}
R.~Brun and F.~Carminati, CERN Program Library Long Writeup W5013 (1993).

\bibitem{zhxsec}
J.~Baglio and A.~Djouadi, J. High Energy Phys. {\bf 10} (2010) 064; 
O.~Brein, R.V.~Harlander, M.~Weisemann, and T.~Zirke, Eur. Phys. J. C {\bf 72}, 1868 (2012).

\bibitem{mcfm}
J.M.~Campbell and R.K.~Ellis, Phys. Rev. D {\bf 60}, 113006 (1999);
{\bf 62}, 114012 (2000);
{\bf 65}, 113007 (2002);
J.M.~Campbell, R.K.~Ellis and C.~Williams, http://mcfm.fnal.gov/.

\bibitem{ttbarxsec}
U.~Langenfeld, S.~Moch, and P.~Uwer, Phys. Rev. D {\bf 80}, 054009 (2009).

\bibitem{dyxsec}
R.~Hamberg, W.L.~van Neerven, and W.B.~Kilgore, Nucl. Phys. {\bf B359},
343 (1991), {\bf B644}; 403(E) (2002).

\bibitem{zptrw} 
V.M.~Abazov {\sl et al.} (D0 Collaboration), Phys. Rev. Lett. {\bf 100}, 102002 (2008).

\bibitem{jetescale} 
V.M.~Abazov {\sl et al.} (D0 Collaboration), Phys. Rev. D {\bf 85}, 052006 (2012).

\bibitem{zjets} 
V.M.~Abazov {\sl et al.} (D0 Collaboration), Phys. Lett. B {\bf 669}, 278 (2008).

\bibitem{sherpa}
T.~Gleisberg, S.~H\"{o}che, F.`Krauss, A.~Sch\"{a}licke, S.~Schumann, and J.-C.~Winter, 
J. High Energy Phys. {\bf 02} (2004) 056;
J.~Alwall {\sl et al.}, Eur. Phys. J. C {\bf 53}, 473 (2008).


\bibitem{dtree}
L.~Breiman, Machine Learning {\bf 45}, 5 (2001).

\bibitem{tmva}
H. Voss {\it et. al.}, Proc. Sci., ({\bf ACAT2007}) (2007) 040 \\
arXiv:physics/0703039.

\bibitem{pdf}
D.~Stump, J.~Huston, J.~Pumplin, W.-K.~Tung, H.-L.~Lai, S.~Kuhlmann, and J.F.~Owens, 
J. High Energy Phys. {\bf 10} (2003) 046.

\bibitem{cls}
T.~Junk, Nucl. Instrum. Methods Phys. Res. A {\bf 434}, 435 (1999);
A.~Read. J. Phys. G {\bf 28}, 2693 (2002).

\bibitem{wade}
W.~Fisher, Report No. FERMILAB-TM-2386-E, 2007.

\bibitem{d0zzwz}
V.M.~Abazov {\sl et al.} (D0 Collaboration), Phys. Rev. D {\bf 85}, 112005 (2012).

\end{thebibliography}
